\documentclass[useAMS,usenatbib]{mn2e}
\usepackage{graphicx,pictex,rotating,color}
\usepackage{txfonts}
\title[Evolution of Class\,0 protostars: Models vs. Observations]{ Evolution of
Class\,0 protostars: Models vs. Observations}

\author[Froebrich et al.]{D. Froebrich$^{1}$\thanks{E-mail: df@cp.dias.ie}, S.
Schmeja$^{2}$, M.D. Smith$^{3, 4}$ and R.S. Klessen$^{2}$\\ $^{1}$Dublin
Institute for Advanced Studies, 5 Merrion Square, Dublin 2, Ireland\\
$^{2}$Astrophysikalisches Institut Potsdam, An der Sternwarte 16, D-14482
Potsdam, Germany\\ $^{3}$Armagh Observatory, College Hill, Armagh BT61 9DG,
Northern Ireland\\ $^{4}$Centre for Astrophysics \& Planetary Science, The
University of Kent, Canterbury CT2 7NR, UK}

\begin{document}

\date{Received sooner; accepted later}
\pagerange{\pageref{firstpage}--\pageref{lastpage}} \pubyear{2005}
\maketitle

\label{firstpage}

\begin{abstract}

The rates at which mass accumulates into protostellar cores can now be 
predicted in numerical simulations. Our purpose here is to develop methods to
compare the statistical properties of the predicted protostars with the
observable parameters. This requires (1) an evolutionary scheme to convert
numerically-derived mass accretion rates into evolutionary tracks and (2) a
technique to compare the tracks to the observed statistics of protostars. 
Here, we use a 3D-Kolmogorov-Smirnov test to quantitatively compare model
evolutionary tracks and observations of Class\,0 protostars.

We find that the wide range of accretion functions and timescales associated
with gravoturbulent simulations naturally overcome difficulties associated with
schemes that use a fixed accretion pattern. This implies that the location of a
protostar on an evolutionary track does not precisely determine the present age
or final accrued mass. Rather, we find that predictions of the final mass for
protostars from observed T$_{\rm bol}$-L$_{\rm bol}$ values are uncertain by a
factor of two and that the bolometric temperature is not always a reliable
measure of the evolutionary stage. Furthermore, we constrain several parameters
of the evolutionary scheme and estimate a lifetime of Class\,0 sources of
2-6\,$\cdot$\,10$^4$\,yrs, which is related to the local free-fall time and thus
to the local density at the onset of the collapse. Models with Mach numbers
smaller than six are found to best explain the observational data. Generally,
only a probability of 70\,\% was found that our models explain the current
observations. This is caused by not well understood selection effects in the
observational sample and the simplified assumptions in the models.

\end{abstract}


\section{Introduction}

A major challenge still remaining in astronomy is to determine how mass
accumulates to form stars. We now know from observations that almost the entire
mass of a star like the Sun is gained during a short protostellar stage (Adams
et al. \cite{1987ApJ...312..788A}). Some of these protostellar objects are
luminous yet cool, suggesting the existence of an even briefer phase within
which the protostar accretes rapidly from a highly obscuring core (Andr\'e et
al. \cite{1993ApJ...406..122A}). The existence of this so-called Class\,0 phase,
however, highlights the need for the development of models which predict how a
protostar abruptly evolves. A successful model would also predict the total
accumulated mass or `final mass', providing a fundamental input for the
subsequent stellar evolution. However, it is debatable whether the development
of low mass stars can be described by an evolutionary model or if the systems
are simply too complex and diverse.

To further the debate, we require three factors: reliable observations of
protostars, models for how cores form and collapse out of a molecular cloud and
a means of converting model parameters into observational parameters. Here, we
develop a statistical method for comparing observational data to model data and
proceed to test the method by making a specific comparison. This requires us to
develop an appropriate technique to compare data within a three dimensional
parameter space. We first introduce, in turn, the data, the model and the
interfacing scheme.

The protostellar stage begins soon after the start of collapse of the cloud
core. Half of the final stellar mass is gained in the very earliest, deeply
embedded phase. Further accretion occurs during the Class\,1 phase while an
accretion disk still feeds material in the extended Class\,2 stage. The number
of observable Class\,0 sources at any given instant is relatively small, due to
their brief lifespans. In addition they are difficult to detect since their
massive envelopes cause their spectral energy distributions (SED) to peak in the
far infrared. Consequently, a variety of strategies have been adopted to
discover Class 0 sources: (1) Large scale imaging for outflows in the optical or
near infrared (NIR; e.g. Ziener \& Eisl\"offel \cite{1999A&A...347..565Z};
Stanke et al. \cite{2002A&A...392..239S}); (2) High resolution processing of
IRAS data (e.g. Hurt \& Barsony \cite{1996ApJ...460L..45H}; O'Linger et al.
\cite{1999ApJ...515..696O}); (3) sub-millimetre or millimetre continuum mapping
(e.g. Shirley et al. \cite{2000ApJS..131..249S}; Motte \& Andr\'e
\cite{2001A&A...365..440M}); (4) deep radio continuum or line surveys (e.g.
Bontemps et al. \cite{1995A&A...297...98B}; Bourke et al.
\cite{1997ApJ...476..781B}). Therefore it is evident to keep in mind that source
samples are subject to strong selection effects.

We require a statistically significant observational sample of protostars with
accurate individual properties. The latter demand a uniform observational
coverage of the entire SED from the NIR to the millimetre range. Froebrich
\cite{2005ApJS..156..169F} recently presented a catalogue with all known
confirmed and candidate Class\,0 sources in the literature. This catalogue
contains 50 objects that possess sufficient observational data to compute the
three main source properties (bolometric temperature and luminosity, envelope
mass; T$_{\rm bol}$, L$_{\rm bol}$, M$_{\rm env}$) accurately.

We can potentially test a range of cloud models. Each model assumes a particular
environment, with initial conditions, boundary conditions, external forces and
feedback effects to be considered (Mac Low \& Klessen
\cite{2004RvMP...76..125M}). However, in the present study, our attention is
focused on a set of accretion rates obtained from numerical simulations of
turbulent clouds (see \S\,\ref{gravomod}; Schmeja \& Klessen
\cite{2004A&A...419..405S}; hereafter SK04). The mass distribution of the cores
so formed can be compared to the initial mass function and the total mass
fraction accumulated in protostellar cores can be related to the overall
star-formation efficiency (Klessen \cite{2001ApJ...556..837K}, Padoan \&
Nordlund \cite{2002ApJ...576..870P}). These recent numerical studies are able to
provide an almost complete model of star formation in clusters, including the
formation of disks and binary systems (see also Bate et al.
\cite{2003MNRAS.339..577B}). As a more fundamental test of these models, we need
to compare model predictions to the observable properties of stars caught in the
act of forming (T$_{\rm bol}$, L$_{\rm bol}$, M$_{\rm env}$). The numerical
models used here, as well as the works from other groups (e.g. Bate et al.
\cite{2003MNRAS.339..577B}) are only able to follow the density and velocity of
material. They are hence very useful in obtaining final mass spectra, binary
fraction, etc.. However, they are not able to predict observables like
bolometric temperature and/or luminosity. To derive those quantities, in
principle a 3D radiative transfer calculation needs to be performed at each time
step. Since this is much too time consuming we applied an evolutionary scheme to
obtain those quantities (see Sect.\,\ref{evolmodel}).

Alternatively, mass accretion rates can be calculated from theoretical and
semi-empirical approaches. In the so-called ``standard solution'' of the
collapse of an isothermal sphere (Shu \cite{1977ApJ...214..488S}) the mass
accretion rate is constant at a value of 0.975\,$c_s^3$\,$/$\,G, where $c_s$ is
the sound speed and G the gravitational constant. Larger but also constant
accretion rates (47\,$c_s^3$\,$/$\,G) are found by Larson
\cite{1969MNRAS.145..271L} and Penston \cite{1969MNRAS.145..457P} for the
collapse of initially uniform isothermal spheres. Henriksen et al.
\cite{1997A&A...323..549H} find that mass ejection and mass accretion both
decline significantly with time during protostellar evolution. Using a
Plummer-type density profile Whitworth \& Ward-Thompson
\cite{2001ApJ...547..317W} predict an accretion rate that possesses a very early
peak during the Class\,0 phase and falls off rapidly after the object develops
into a Class\,1 protostar. Shu et al. \cite{2004ApJ...601..930S} obtained
similar results through analytical and numerical works. Hydrodynamical
simulations of star formation (Foster \& Chevalier \cite{1993ApJ...416..303F};
Tomisaka \cite{1996PASJ...48L..97T}) controlled by supersonic turbulence (e.g.
Klessen \cite{2001ApJ...550L..77K}; SK04) yield similar behaviour.

In theory, there are two means to interface model and observed data sets. Ideal
would be to derive a set of mass accretion rates directly from observations of
protostellar cores but infall speeds are extremely difficult to measure. Hence,
conversely, we convert the mass infall rates from the models into the observable
quantities. The main three are the bolometric temperature and luminosity,
T$_{\rm bol}$ and L$_{\rm bol}$, and the envelope mass, M$_{\rm env}$.

Several such conversion schemes have been developed. Evolutionary models were
first discussed by Bontemps et al. \cite{1996A&A...311..858B} and Saraceno et
al. \cite{1996A&A...309..827S}. Along these lines, Andr\'e et al.
\cite{2000prpl.conf...59A} took an exponentially decline in both envelope mass
and accretion rate to predict L$_{\rm bol}$-M$_{\rm env}$ tracks. An analytical
scheme was presented by Myers et al. \cite{1998ApJ...492..703M}. They assumed a
mass infall rate matching the Shu solution at early times and then displaying an
exponential fall off with time. The core which supplies the mass also provides
the obscuration, thus fixing evolutionary tracks on a T$_{\rm bol}$-L$_{\rm
bol}$  diagram. Based on this work, Smith
(\cite{1998Ap&SS.261..169S,2000IrAJ...27...25S, s02}; S98 hereafter) presented
an evolutionary scheme but adopted alternative analytical forms for the mass
accretion (exponential increase and power law decrease) which resemble the
results obtained by Whitworth \& Ward-Thompson \cite{2001ApJ...547..317W} or
SK04. This evolutionary model, founded on mass transfer between the envelope,
disc, protostar and jets, has provided successful interpretations for various
sets of observational data (T$_{\rm bol}$, L$_{\rm bol}$, M$_{\rm env}$, outflow
luminosity) of Class\,0 and Class\,1 objects (e.g. Davis et al.
\cite{1998MNRAS.299..825D}; Yu et al. \cite{2000AJ....120.1974Y}; Stanke
\cite{2000PhDT........12S}; Froebrich et al. \cite{2003MNRAS.346..163F}). Here,
we adopt the S98 scheme but replace the analytical accretion rates with those
derived numerically. We note that this scheme does not include any inclination
effects that become very important during the Class\,1 phase for T$_{\rm bol}$
and L$_{\rm bol}$ determinations. We thus focus our work on Class\,0 objects.

Our approach in this paper is as follows. We combine different sets of mass
accretion rates obtained from gravoturbulent molecular cloud fragmentation
calculations (gt-models; Klessen et al. \cite{2000ApJ...535..887K}; Klessen
\cite{2001ApJ...550L..77K}; Appendix\,\ref{details_gtmodels}), as described in
detail in SK04, and the evolutionary scheme (e-model) from S98 (see also
Appendix\,\ref{details_emod}). For each set of mass accretion rates, we
determine evolutionary tracks in the T$_{\rm bol}$-L$_{\rm bol}$-M$_{\rm env}$
parameter space and compare them with the distribution of the observed Class\,0
sample. A 3D-Kolmogorov-Smirnov test (KS-test, see also
Appendix\,\ref{details_kstest}) yields a probability for the two distributions
to be drawn from the same basic population (called {\it agreement} hereafter).
The agreement parameter will thus determine how well a model is able to explain
the currently available set of observations. 

We remark that two separate models (gt-models and e-models) are needed to
obtain the model protostellar quantities that can be compared to the
observational data. Thus, we are only able to determine how well the {\it
combination} of {\it both} can explain the observations. We are not able to
decide which model is responsible for potential disagreement with the
observations. However, the variation of free parameters in the models and a
search for correlations of the parameter values with the agreement will help to
define constraints and provide suggestions for improved model combinations.

In Sect.\,\ref{methods} we describe the observational data and our models.
Results and the discussion are put forward in Sect.\,\ref{analysis}. Details of
the gravoturbulent and evolutionary models and the Kolmogorov-Smirnov test are
described in Appendixes\,\ref{details_gtmodels} to \ref{details_kstest}.

\section{Observations and Models}
\label{methods}

\subsection{Observational data}
\label{obsdata}

Several samples of Class\,0 sources have been published (e.g. Chen et al.
\cite{1995ApJ...445..377C}; Andr\'e et al. \cite{2000prpl.conf...59A}; Shirley
et al. \cite{2000ApJS..131..249S}; Motte \& Andr\'{e}
\cite{2001A&A...365..440M}). The latest study, which combines all previous ones
and computes the object properties uniformly from all published data, is
presented by Froebrich \cite{2005ApJS..156..169F}. There, a list of 95 confirmed
or candidate objects was compiled. Fifty of these sources possess sufficient
observational data to be classified as Class\,0 or Class\,0/1 objects and allow
us to determine the three source properties T$_{\rm bol}$, L$_{\rm bol}$, and
M$_{\rm env}$ accurately. To ensure a reliable comparison of models and
observations, we apply the following restrictions to this sample:

(1) We omit sources that have distances larger than 500\,pc. This reduces the
bias towards higher mass objects.

(2) All objects in Taurus are underluminous compared to the other sources,
considering T$_{\rm bol}$ and M$_{\rm env}$. Note that this does not only mean a
low luminosity for these objects, but rather a low luminosity combined with a
high envelope mass. The three very low luminosity objects in our sample (see
Fig.\,\ref{tracks}) combine low luminosity (1\,L$_\odot$) with low envelope mass
(0.2\,M$_\odot$) and might form (very) low mass stars. In total, 25\,\% of the
Class\,0 objects in the sample are underluminous. We exclude them for the
comparison between the models and observations (but see the discussion in
Sect.\,\ref{underlumi}). 

(3) A histogram of the number of sources in a certain T$_{\rm bol}$-bin shows
that the observational sample is very incomplete at high bolometric
temperatures. Hence the comparison of observations and models is restricted to
T$_{\rm bol} <$\,80\,K.

This selection procedure leaves us with a sample of 27 Class\,0 sources. This
sample consists mostly of objects in Perseus, Orion and Serpens.

\subsection{Gravoturbulent models}
\label{gravomod}

We performed numerical simulations of the fragmentation and collapse of
turbulent, self-gravitating gas clouds and the resulting formation and evolution
of protostars. We employed a code based on smoothed particle hydrodynamics (SPH;
Monaghan \cite{1992ARA&A..30..543M}) in order to resolve large density contrasts
and to follow the evolution over a long timescale. The code includes periodic
boundary conditions (Klessen \cite{1997MNRAS.292...11K}) and sink particles
(Bate et al. \cite{1995MNRAS.277..362B}) that replace high-density cores while
keeping track of mass, linear and angular momentum. A detailed study of the 24
sets of protostellar mass accretion rates resulting from the gt-models is
presented in SK04 (see also Jappsen \& Klessen \cite{2004A&A...423....1J}) and
we adopt their gt-model notation.

We now define when a model core is considered to be a Class\,0 object.
Observationally, a protostar is a Class\,0 source according to the definition
given in Andr\'e et al. \cite{2000prpl.conf...59A}. It is shown by Andr\'e et
al. \cite{1993ApJ...406..122A} that this definition is roughly equivalent to the
physical state where the mass in the envelope exceeds the mass of the central
protostar. Thus, a model protostar is considered a Class\,0 object when (a) the
mass of the central core exceeds 10$^{-2}$M$_\odot$ (this corresponds to the
time when the second hydrostatic core is formed in the centre (Larson
\cite{2003RPPh...66.1651L}) and distinguishes the Class\,0 sources from the
pre-stellar core phase) and (b) the mass of the envelope is larger than the mass
of the central star. This separates the Class\,0 from the Class\,1 objects. 

Small number statistics will always be a major concern when comparing models
and observations of Class\,0 objects. In order not to introduce further
uncertainties from the model side, we restricted our analysis to those
gt-models that possess more than 37 stars and have a numerical resolution of at
least 2$\cdot$10$^5$ particles. This leaves 16 out of the 24 sets of accretion
rates of SK04 for further analysis. In addition, accretion rates for individual
protostars from the gt-models are smoothed by a viscous time scale of the
accretion disc of $\sim 10^4$\,yrs (see Appendix\,\ref{details_gtmodels} for
details).

A small fraction of model protostars were highly accelerated (e.g. by ejection
from a multiple system). Due to the adopted periodic boundary conditions in our
calculations, these objects cross the computational domain many times while
continuing to accrete. In reality, however, these protostars would have quickly
left the high-density gas of the star-forming region and would not be able to
gain more mass. We therefore consider accretion to stop after the object has
crossed the computational box more than ten times. 

The radius of a sink particle is fixed at 280\,AU, the physics (e.g. exact
accretion, radiation) inside this volume cannot be resolved. Therefore, an
evolutionary model describing the processes inside the sink particle is
required.

\begin{table}
\begin{center}
\caption{\label{params} Variable parameters from the e-model of S98, together
with the range they were varied in and the original value. The last two columns
provide the overall range found to lead to the best agreement.}
\begin{tabular}{lccccc}
Parameter & \multicolumn{2}{c}{varied} & org. value & \multicolumn{2}{c}{best} \\
 & from & to & & from & to \\
\noalign{\smallskip}
\hline
\noalign{\smallskip}
$T_{\rm env}$\,[K] & 10 & 30 & 24 & 15 & 19 \\
$frac_{\rm env}$\,[\%] & 75 & 100 & 87 & 86 & 96 \\
$M_{\rm extra}$ & 0 & 5 & 2 & 1.0 & 2.4 \\
$t_0$\,[10$^3$\,yrs] & 1 & 100 & 30 & 30 & 80 \\
$\alpha$ & 0.5 & 4.0 & 1.75 & 1.4 & 3.2 \\
$R_{\rm in}$\,[AU] & 5 & 100 & 30 & 35 & 80 \\
$M_{\rm eff}$\,[\%] & 0 & 50 & 30 & 36 & 44 \\
$p$ & 1.4 & 2.0 & 1.5 & 1.55 & 1.80 \\
$\kappa$\,[cm$^2$g$^{-1}$] & 2 & 6 & 4 & 3.0 & 5.0 \\
\end{tabular}
\end{center}
\end{table}

\subsection{Evolutionary scheme}
\label{evolmodel}
 
The e-model is used to transform the mass accretion rates from the gt-models
into observable quantities such as T$_{\rm bol}$, L$_{\rm bol}$, and M$_{\rm
env}$. It is based on mass and energy transfer between the different components
of the forming star (protostar, disc, envelope, jet). Mass conservation is the
main principle in this analytical model. Each simulated accretion rate from the
gt-models is taken as the mass inflow rate from a spherical envelope onto the
inner disc/protostar/jet system. All the mass flows through the disc but only a
fraction accretes onto the protostar. The rest is ejected within two jets. The
ejected mass fraction is small but not negligible in the Class\,0 stage. See
Appendix\,\ref{details_emod} for a detailed description of the parameters and
equations of the evolutionary model used in this work.

Many parameters and constants are needed to fully describe the envelope and
predict the radiative properties (Myers et al. \cite{1998ApJ...492..703M}, S98,
Appendix\,\ref{details_emod}). We carefully chose a subset of parameters which
we kept variable: (1) $T_{\rm env}$; The temperature of the outer envelope, out
to which the envelope mass is determined. (2) $frac_{\rm env}$; The total mass
that will be accreted onto the protostar is distributed in the envelope and the
disc. A constant ratio of these two masses is assumed and $frac_{\rm env}$
represents the fraction of the total mass that is in the envelope. (3) $M_{\rm
extra}$; This is the supplementary mass fraction in the immediate surroundings
which does not actually fall towards the star, to be accreted or jetted away,
but is removed directly from the core probably through a feedback process. See
Eq.\,\ref{eq_mextra} for the exact definition used. (4/5) $t_0$, $\alpha$; These
two parameters describe the dispersion of the extra mass in the envelope. (6)
$R_{\rm in}$; The inner radius of the envelope. (7) $M_{\rm eff}$; The  maximum
percentage of the infalling envelope mass which is ejected into the outflow. (8)
$p$; The power-law index of the density distribution in the envelope, assumed
constant in time. (9) $\kappa$; The opacity of the envelope material at
12\,$\mu$m. In Table\,\ref{params} we list the parameter ranges which were
tested. 

We now only need to impose a mass accretion rate from the gt-models for the
evolution of a model protostar to be fully determined. Using all individual
accretion rates in a gt-model allows us to build up a T$_{\rm
bol}$-L$_{\rm bol}$-M$_{\rm env}$ distribution which can be compared to the
observational dataset.

Some of our free parameters can be, and have been, observed for individual
objects. This includes for example the power-law index of the density
distribution in the envelope (e.g. Chandler \& Richer
\cite{2000ApJ...530..851C}, Motte \& Andr\'e \cite{2001A&A...365..440M}) or
$frac_{\rm env}$ (e.g. Looney et al. \cite{2003ApJ...592..255L}). We chose to
keep those parameters variable in order to test our method, since we expected to
obtain values within the observational constraints.

\begin{figure*}
\includegraphics[width=5.5cm, bb=10 10 515 427]{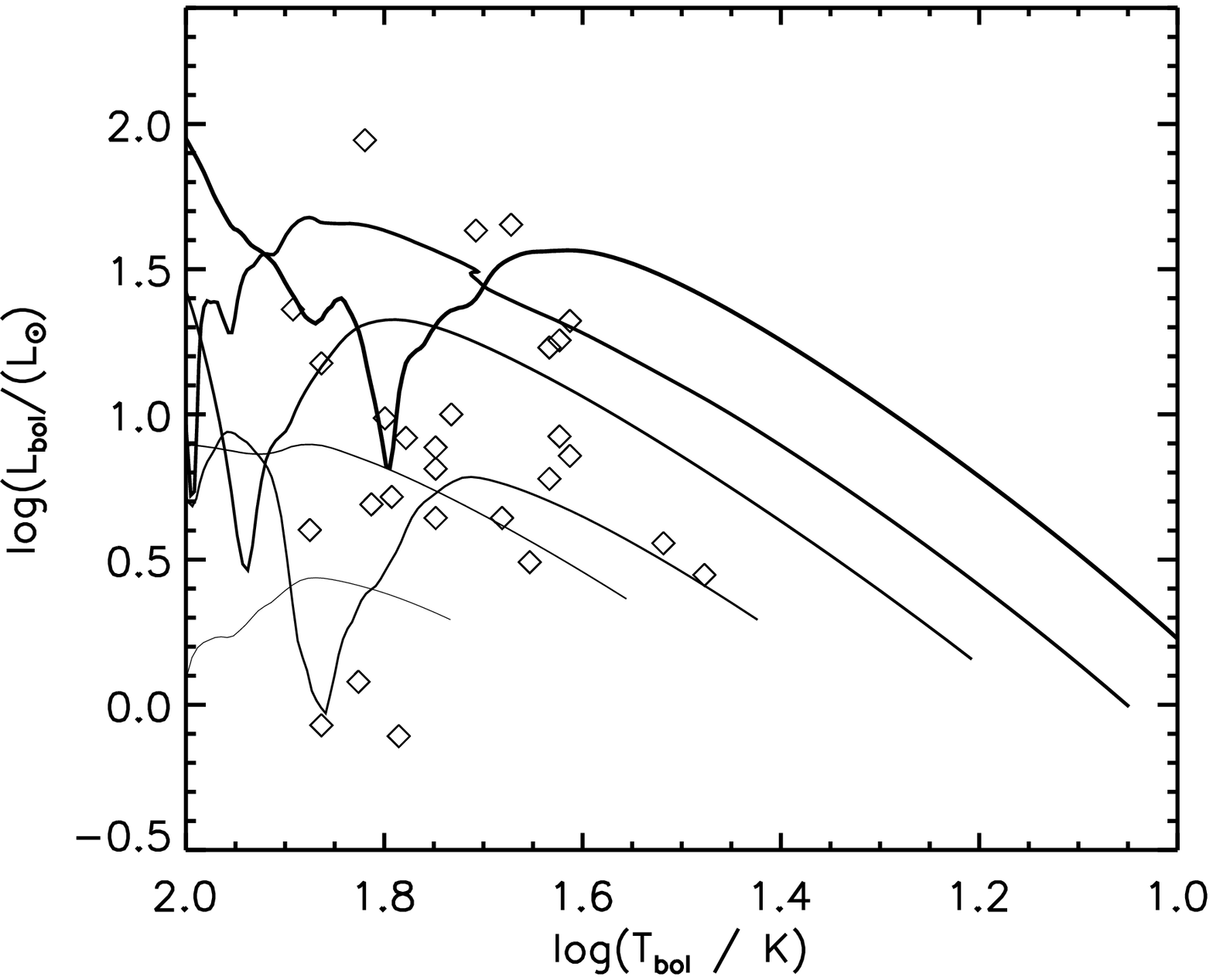} 
\includegraphics[width=5.5cm, bb=10 10 515 427]{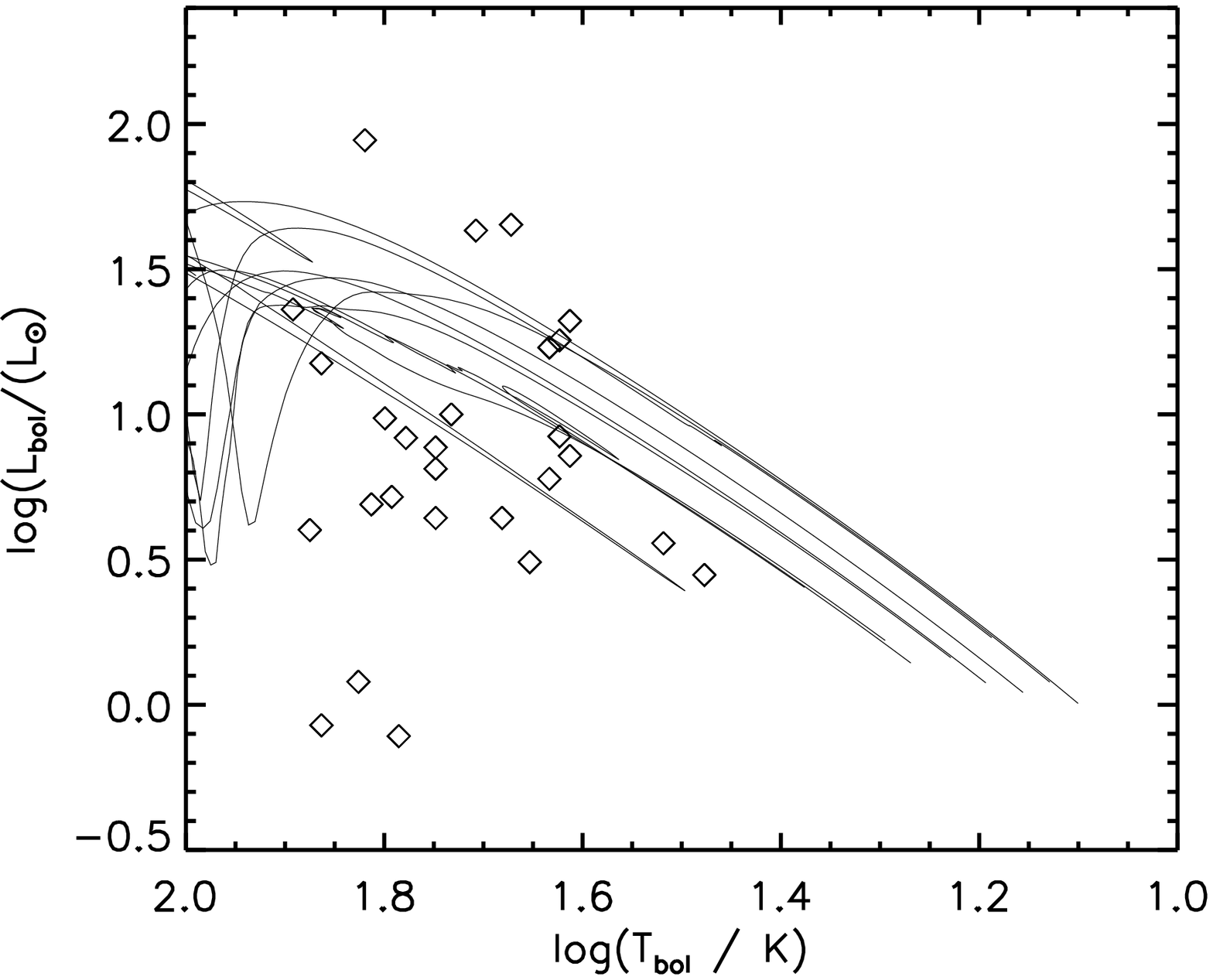} 
\includegraphics[width=5.5cm, bb=10 10 515 427]{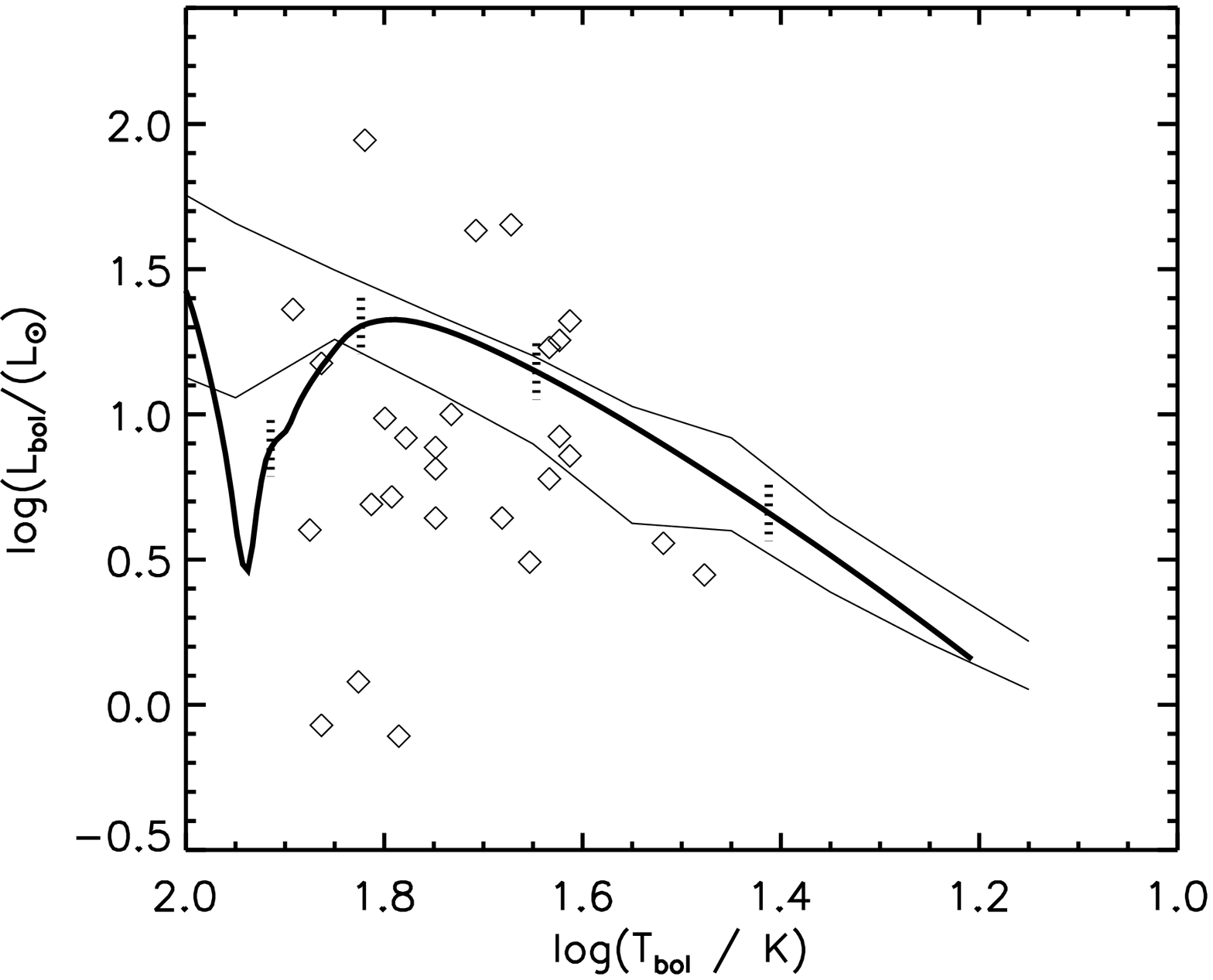} \\
\includegraphics[width=5.5cm, bb=10 10 515 427]{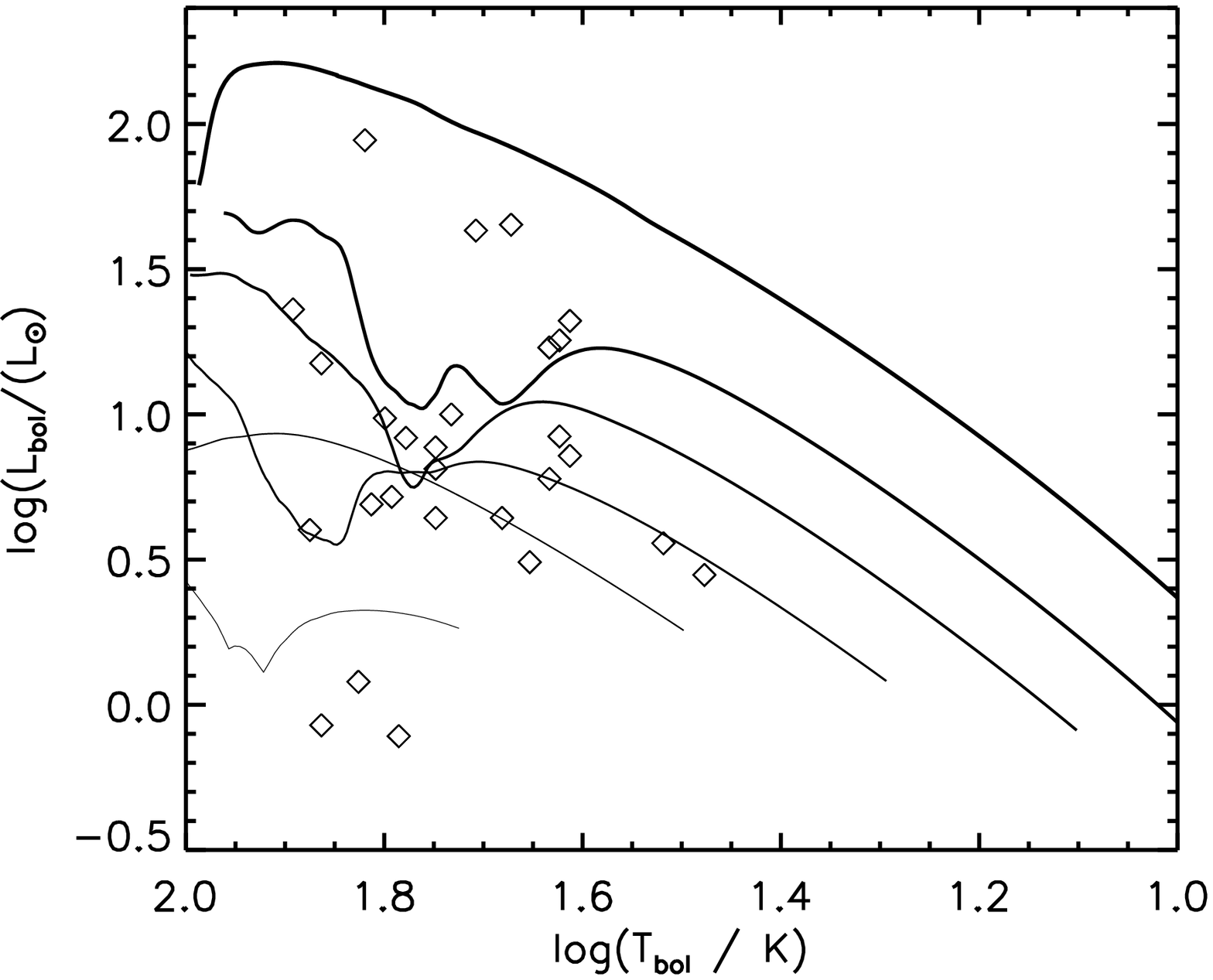} 
\includegraphics[width=5.5cm, bb=10 10 515 427]{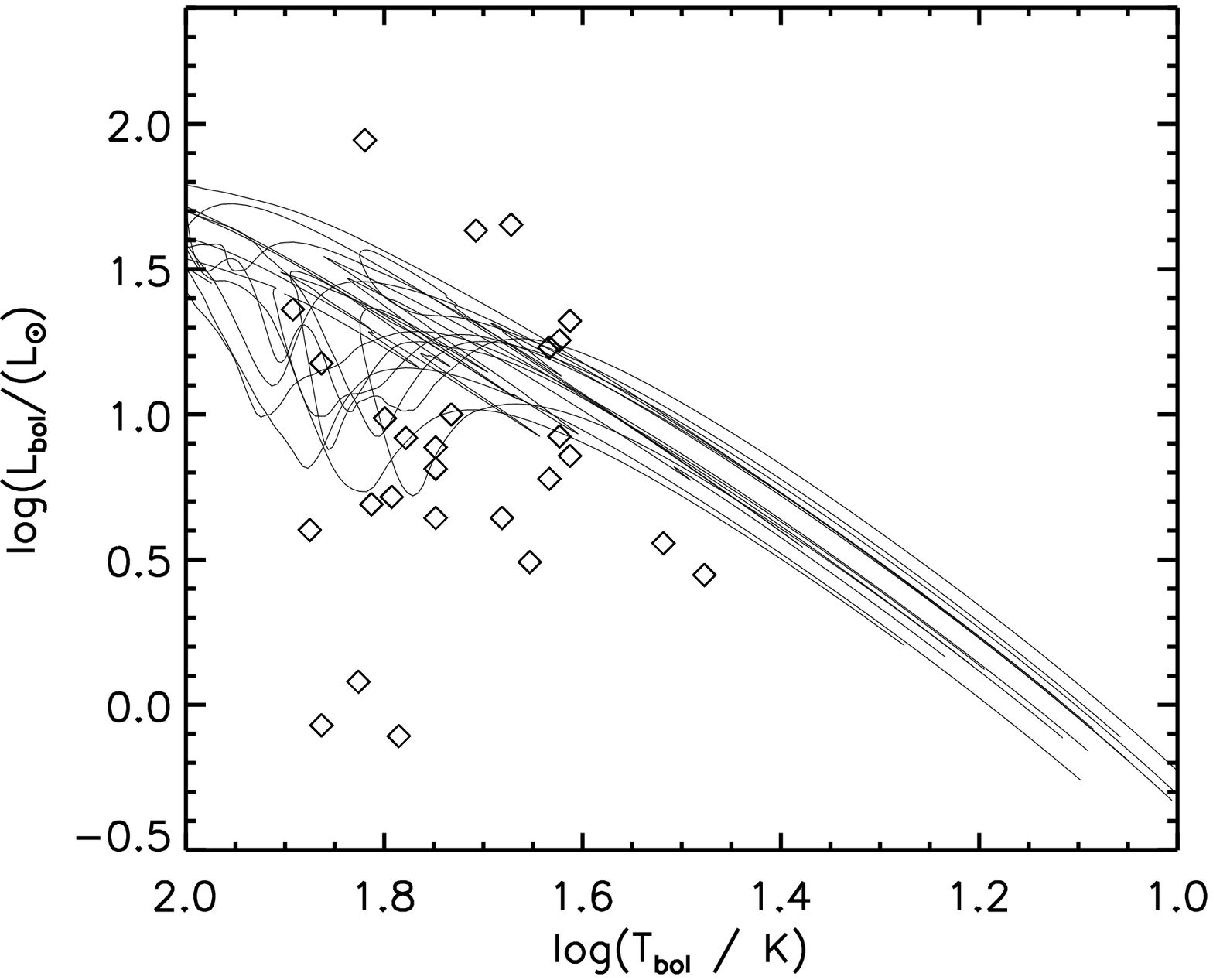} 
\includegraphics[width=5.5cm, bb=10 10 515 427]{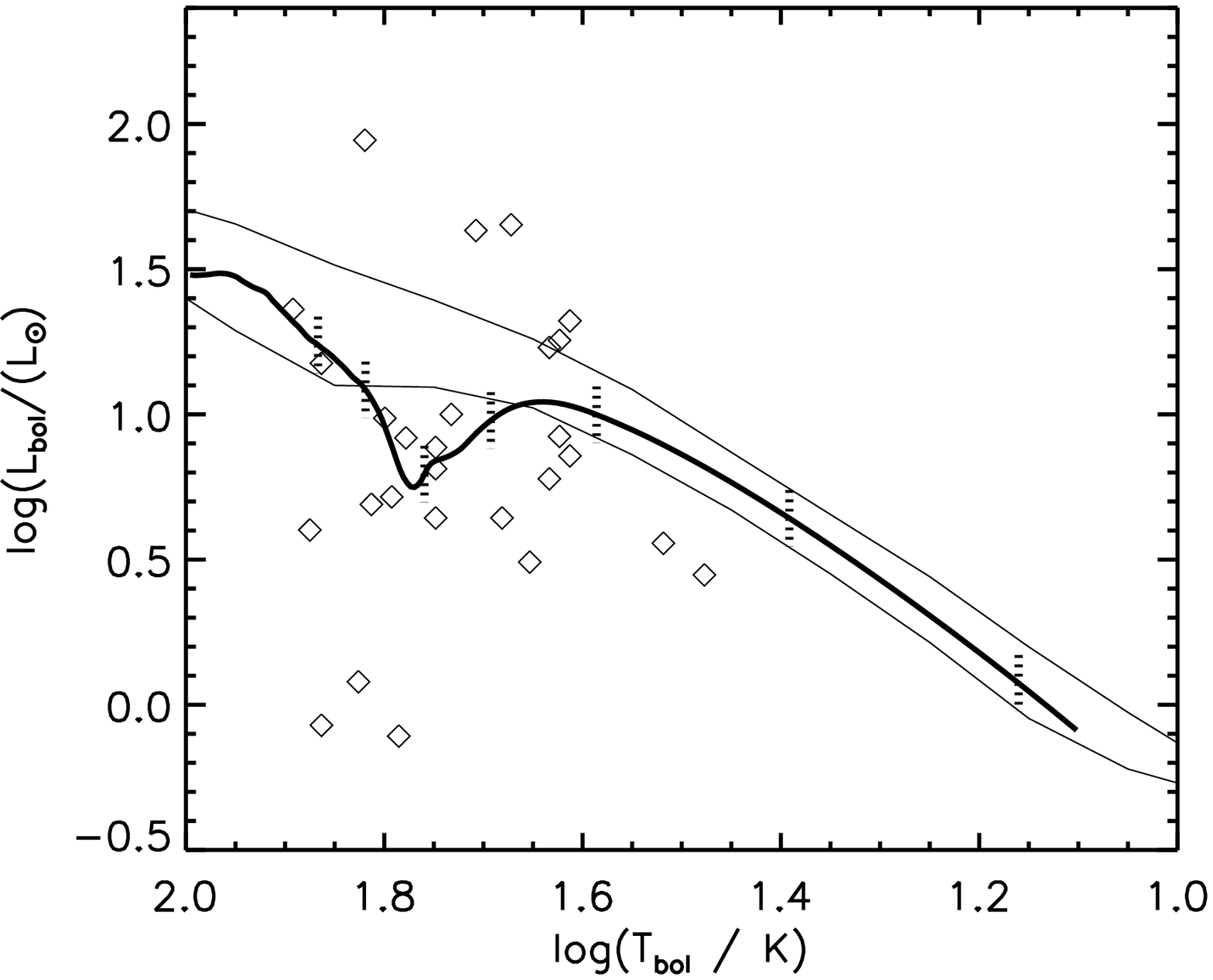} \\
\includegraphics[width=5.5cm, bb=10 10 515 427]{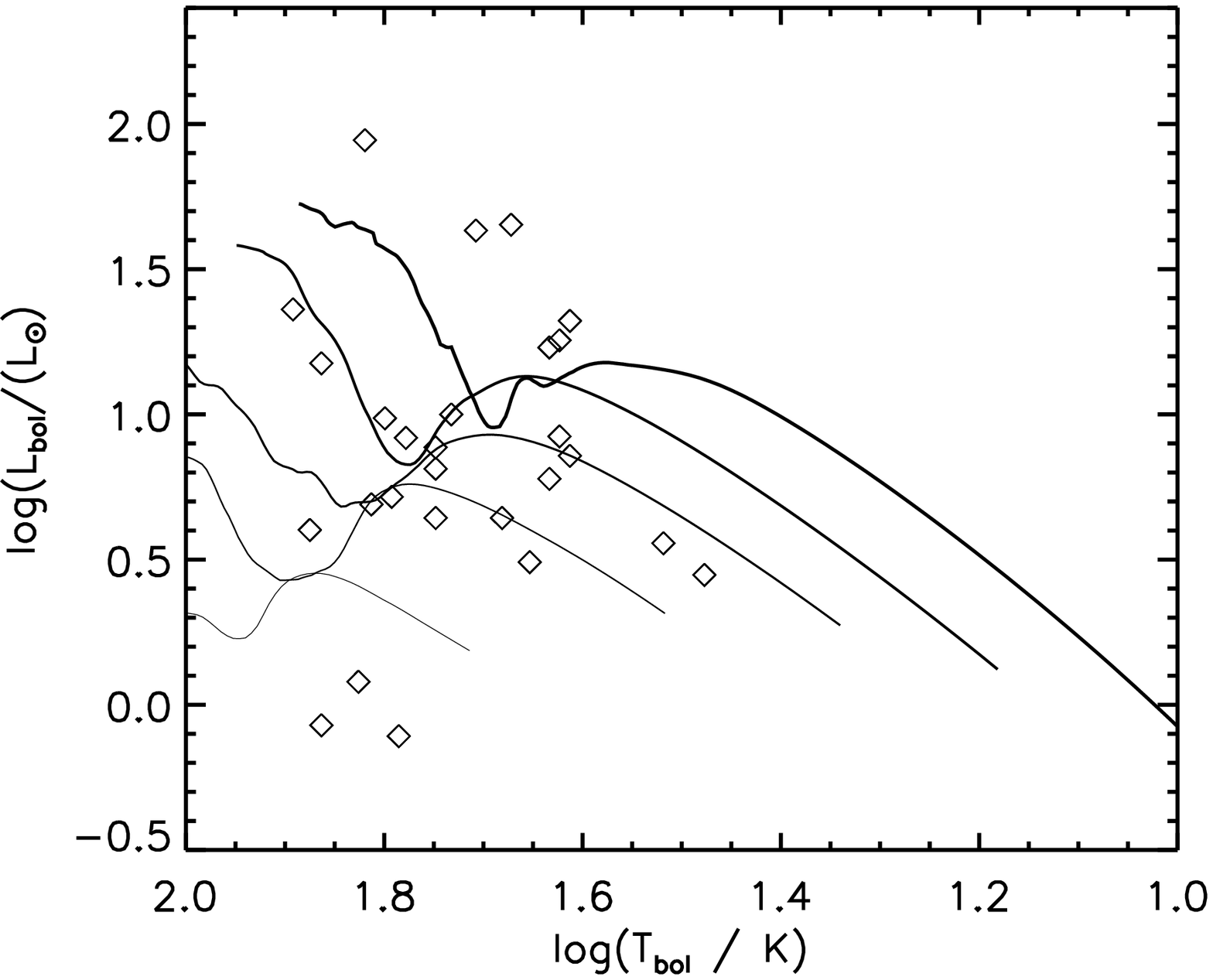} 
\includegraphics[width=5.5cm, bb=10 10 515 427]{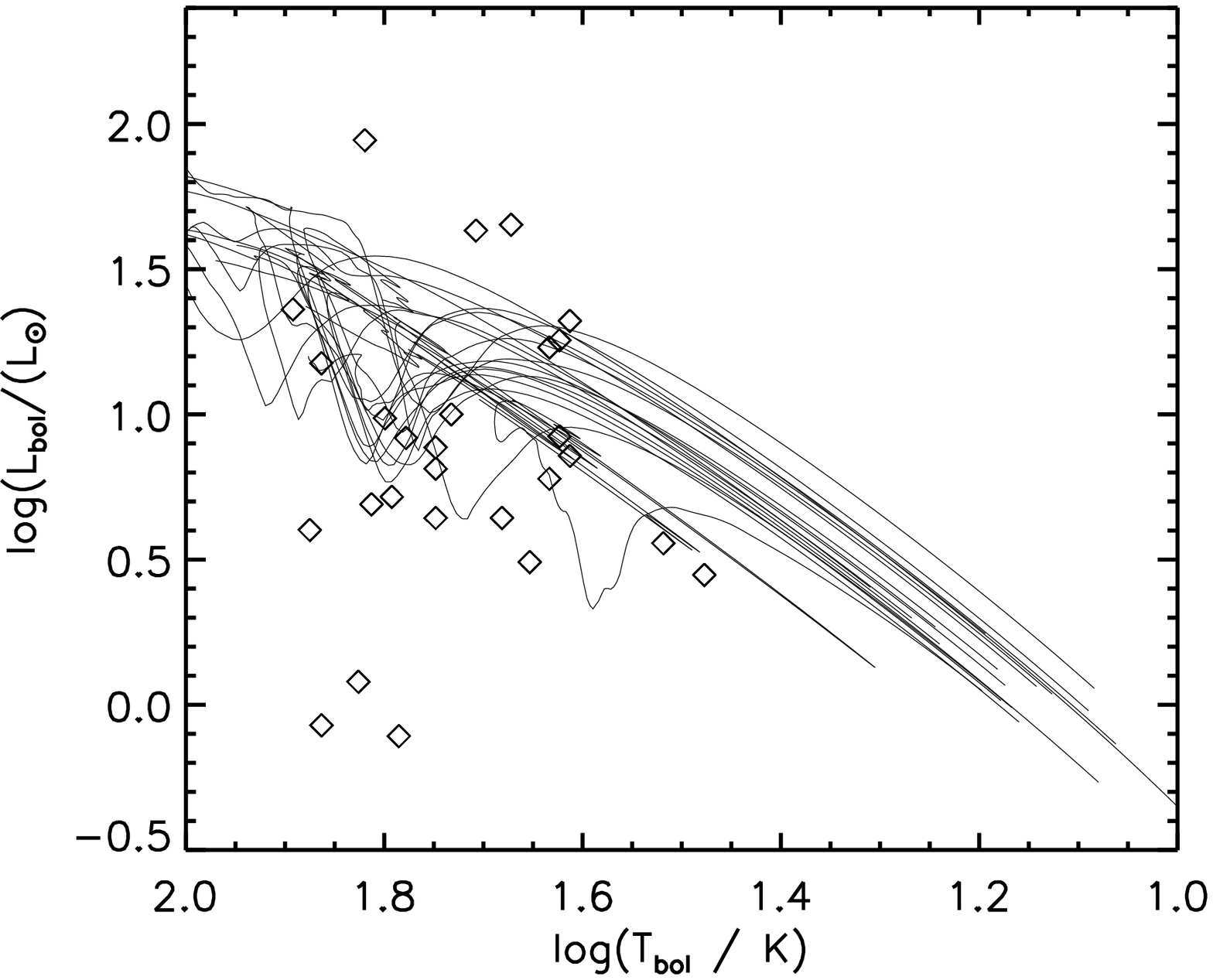} 
\includegraphics[width=5.5cm, bb=10 10 515 427]{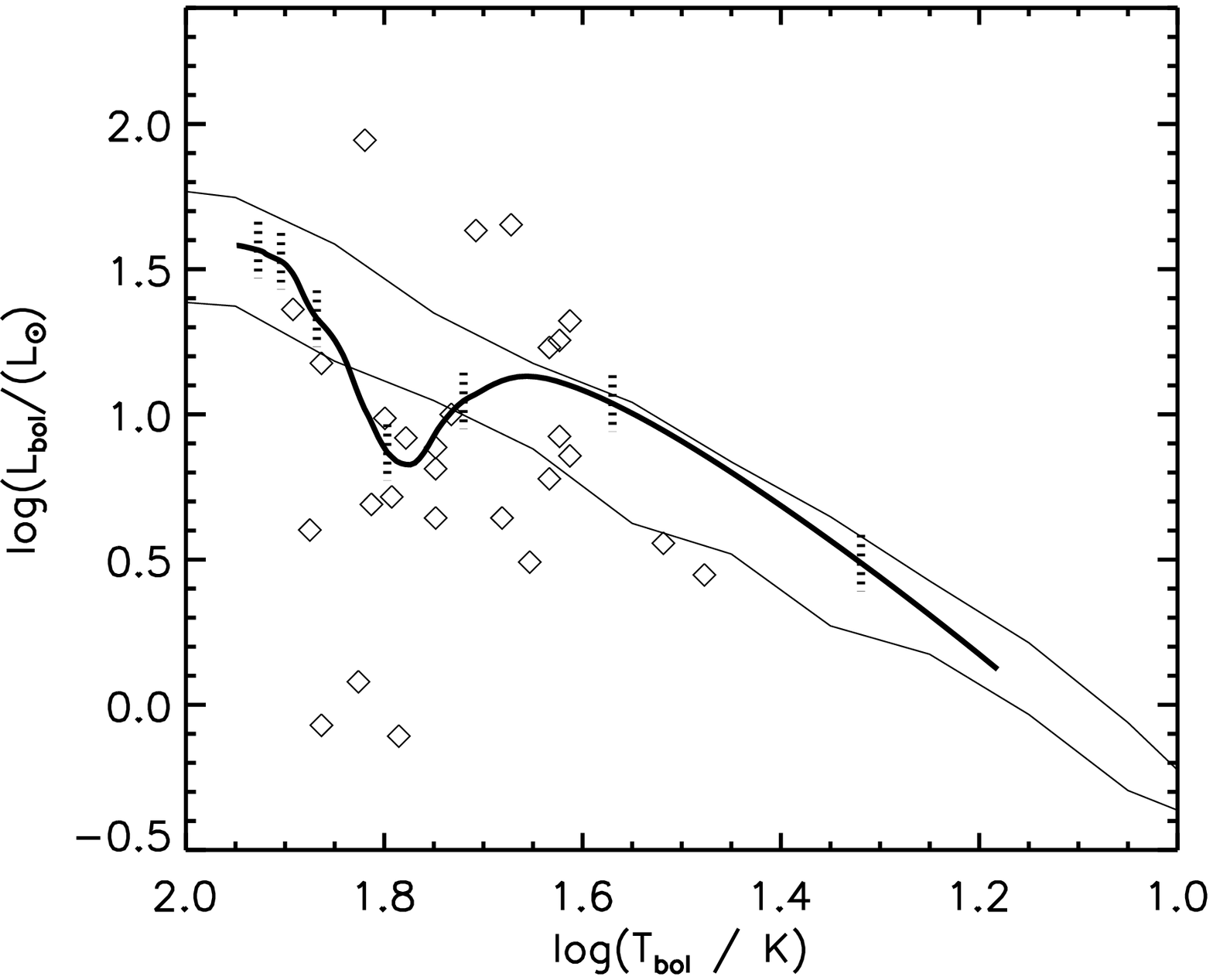} \\
\caption{\label{tracks} Evolutionary tracks in the T$_{\rm bol}$-L$_{\rm bol}$
parameter space of the model Class\,0 sources in the gt-models M05k8 (top row),
M2k2 (middle row), and M6k2a (bottom row). The tracks are determined using
typical e-model parameters from the range best fitting the observations (see
Table\,\ref{best_agree}). Left column: Tracks for the average accretion rates in
the six mass bins. Thicker lines correspond to higher final masses. Note that
there are no stars in mass bin\,6 in the M6k2a model. Middle column: All
individual tracks in mass bin\,4. Right column: Track for average accretion
rate of stars in mass bin\,4 (solar mass stars; thick line) and the
$\pm$\,1\,$\sigma$ scatter of the individual tracks (thin lines). The vertical
lines mark ages of the sources starting from 5\,$\cdot$\,10$^3$\,yrs in steps of
5\,$\cdot$\,10$^3$\,yrs. The open diamonds in each panel mark the positions of
the observational sample (taken from Froebrich 2005). See
Sect.\,\ref{mass_estimate} for more details.}
\end{figure*}

\subsection{3D KS-Test and probabilities}

The best way to compare two distributions of data points is through a
Kolmogorov-Smirnov test. This test yields the probability of two distributions
being drawn from the same basic population. In one dimension, the KS-test
compares the cumulative probability functions of a sample and a model and for
large sample sizes the probability can be determined analytically. Since our
distributions are three dimensional (T$_{\rm bol}$-L$_{\rm bol}$-M$_{\rm env}$),
there are no analytical means to perform such a test. We therefore generalised
the method of a two dimensional KS-test (e.g. Singh et al.
\cite{2004PhRvD..69f3003S}) for our purpose. The basic result of this test
is the value $D_{\rm 3D}$ that ranges from zero to one. The lower this value the
better the two distributions match. Using a Monte Carlo method $D_{\rm 3D}$ can
be converted to the agreement ($P_{\rm 3D}$), which gives the probability that
the two distributions are drawn from the same population. 

The task of finding the parameter set for the e-models that results in the
highest agreement is a multi-dimensional, non-linear, minimisation problem. We
solved this using a Monte Carlo approach. In Appendix\,\ref{details_kstest} we
outline the details of this method which allows to constrain the range for the
e-model parameters and to determine the best agreement.

\section{Analysis \& Discussion}
\label{analysis}

\subsection{Evolutionary tracks}
\label{mass_estimate}

One of our goals is to investigate the accuracy to which the final mass of a
protostar (M$_{\rm final}$) can be estimated from its present location in the
T$_{\rm bol}$-L$_{\rm bol}$ diagram. To achieve this, we sort for each gt-model
the individual model stars into final mass bins with a width of 0.3 in
logarithmic units (equivalent to a factor two in mass). The following ranges for
the mass bins were chosen: $<$\,0.2, 0.2...0.4, 0.4...0.8, 0.8...1.6, 1.6...3.2,
$\ge$\,3.2\,M$_\odot$. The particular size of the mass bins was adopted in order
to ensure a reasonably large number of stars in each bin for the gt-models.
Depending on the mass function and the total number of objects in the gt-model,
there are up to 20 objects per bin for the lower masses (M$_{\rm final}
\le$\,1.6\,M$_\odot$). The higher mass bins naturally suffer from a paucity of
objects. 

We determined for every gt-model the average accretion rate history in each mass
bin. These mean accretion rates are then used to determine evolutionary tracks
in the T$_{\rm bol}$-L$_{\rm bol}$ diagram. The left column of
Fig.\,\ref{tracks} shows the tracks for the six mass bins of the models M05k8,
M2k2, and M6k2a as an example. The lowest final masses correspond to the
thinnest line, and so forth. Note that the average evolutionary tracks show
higher L$_{\rm bol}$ at a given T$_{\rm bol}$ for higher final star masses. This
represents the fact that stars with higher final masses on average have higher
accretion rates (SK04) and hence bolometric luminosities. 

Further we analysed the individual evolutionary tracks. As an example, we plot
all individual tracks for the same three models with final stellar masses in the
range 0.8 to 1.6\,M$_\odot$ (mass bin\,4, solar mass stars) in the middle column
of Fig.\,\ref{tracks}. These tracks are determined assuming typical e-model
parameters from the range that best fit the observations (see below). We find
that all tracks show a similar general behaviour. The tracks start off at
T$_{\rm bol}$\,$\approx$\,15\,K for solar mass stars. Lower mass Class\,0
sources start off at higher temperatures. Then they show a gradual increase in
luminosity until the end of the Class\,0 phase. Some notable exceptions are
evident in Fig.\,\ref{tracks}, which imply that T$_{\rm bol}$ is not always a
reliable guide as to the envelope-protostar mass ratio. 

Are we able to estimate the final mass of a Class\,0 source from its position 
in the T$_{\rm bol}$-L$_{\rm bol}$ diagram? To address this question, we
determined the 1\,$\sigma$ scatter of the individual tracks at each time step in
all mass bins. In the right column of Fig.\,\ref{tracks} this scatter is shown
for mass bin\,4 as thin lines. Almost independent of the gt-model, we find that
the scatter has about the size of the separation between tracks for two adjacent
mass bins. Hence, we conclude that from a certain position in the T$_{\rm
bol}$-L$_{\rm bol}$ diagram, {\em we are only able to estimate the final mass of
the protostar to within a factor of two.} Note that this estimate does not take
into account possible errors in the measurement of T$_{\rm bol}$ and L$_{\rm
bol}$, as well as possible different accretion histories. This supports the
notion that other observables might be more adequate for dating Class\,0
protostars, e.g. the L$_{\rm smm}$/L$_{\rm bol}$ ratio (Young \& Evans
\cite{2005ApJ...627..293Y}), in agreement with the original observational
definition of Class\,0 sources (Andr\'e et al. \cite{1993ApJ...406..122A}).

We further find that tracks determined from averaged accretion rates noticeably
differ from the averaged evolutionary tracks of the individual stars. This is
evident in the right column of Fig.\,\ref{tracks} which shows as a thick line
the track from the averaged accretion rate. At some points, it approaches or
even lies outside the $\pm$\,1\,$\sigma$ range of the individual evolutionary
tracks. This clearly shows, together with the findings in the above paragraph,
that the individual accretion history significantly influences the position in
the T$_{\rm bol}$-L$_{\rm bol}$ diagram, in addition to the final stellar mass.

The general behaviour of the T$_{\rm bol}$-L$_{\rm bol}$ tracks (characteristics
of the individual tracks, the average tracks, and the scatter around the
average) is independent of the considered mass bin and gt-model.

\begin{figure*}
\beginpicture
\setcoordinatesystem units <44.06mm,15.20mm> point at 0 0
\setplotarea x from 1 to 2 , y from -0.5 to 2.4
\put {\includegraphics[width=4.406cm, height=1.52cm, bb=0 157 566 323]{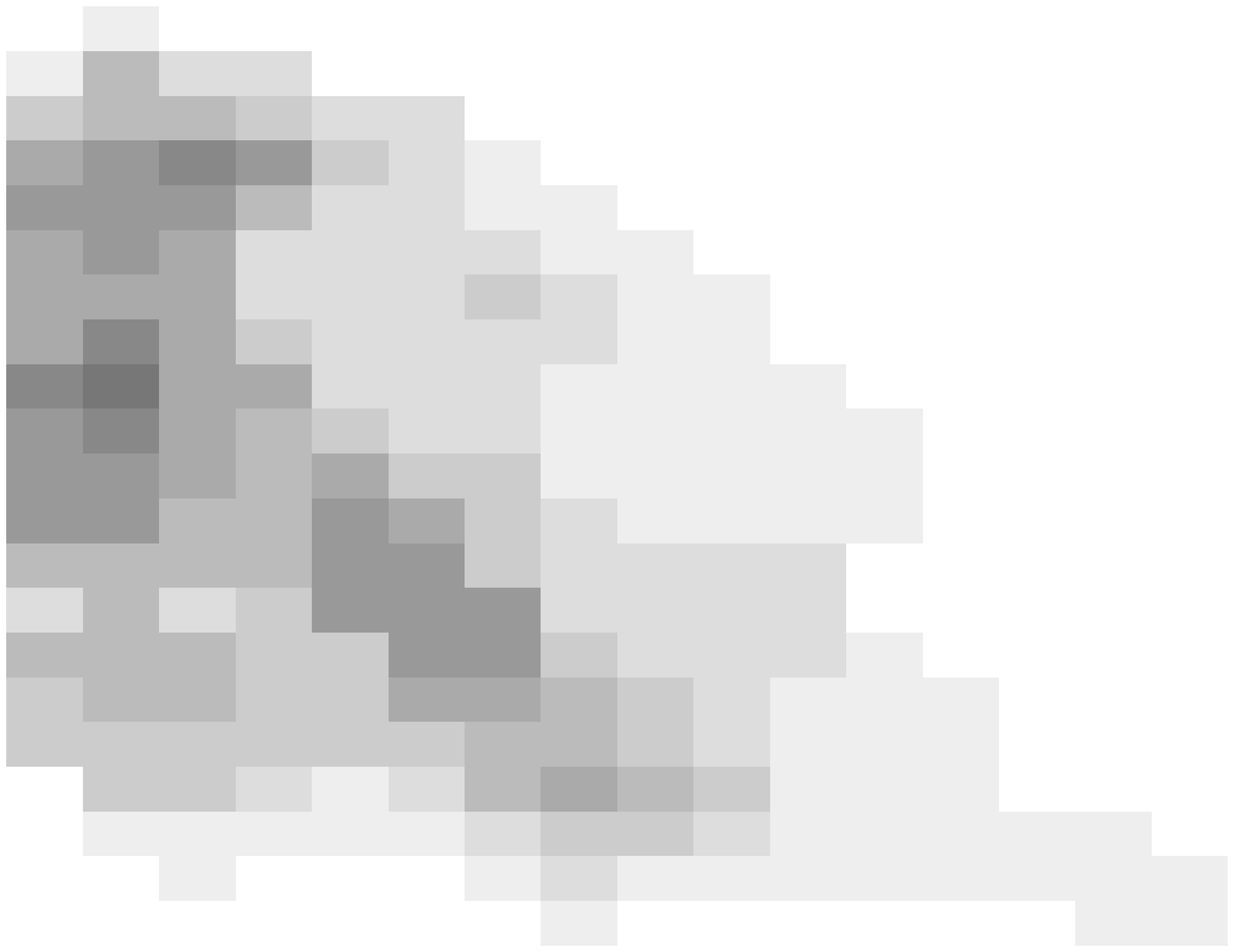}} at 1.5 0.95
\axis left label {}
ticks in long numbered from -0.0 to 2 by 1
      short unlabeled from -0.5 to 2. by 0.5 /
\axis right label {}
ticks in long unlabeled from -0.0 to 2 by 1
      short unlabeled from -0.5 to 2. by 0.5 /
\axis bottom label {}
ticks in long unlabeled from 1 to 2 by 0.2
      short unlabeled from 1 to 2 by 0.1 /
\axis top label {}
ticks in long unlabeled from 1.0 to 2 by 0.2
      short unlabeled from 1 to 2 by 0.1 /
\put {2.0} at 1.0 -0.7
\put {1.8} at 1.2 -0.7
\put {1.6} at 1.4 -0.7
\put {1.4} at 1.6 -0.7
\put {1.2} at 1.8 -0.7
\put {1.0} at 2.0 -0.7
\put {log[T$_{\rm bol}$/K]} at 1.5 -1.
\put {\begin{sideways} log[L$_{\rm bol}$/L$_\odot$] \end{sideways}} at 0.87 0.95
\put {{\color{black}$\circ$}} at  1.21467 -0.10790
\put {{\color{black}$\circ$}} at  1.13668 -0.07058
\put {{\color{black}$\circ$}} at  1.17393  0.07918
\put {{\color{black}$\circ$}} at  1.26761  1.00000
\put {{\color{black}$\circ$}} at  1.13668  1.17609
\put {{\color{black}$\circ$}} at  1.36653  1.23045
\put {{\color{black}$\circ$}} at  1.37675  1.25527
\put {{\color{black}$\circ$}} at  1.52288  0.44715
\put {{\color{black}$\circ$}} at  1.38722  1.32222
\put {{\color{black}$\circ$}} at  1.10791  1.36173
\put {{\color{black}$\circ$}} at  1.34679  0.49136
\put {{\color{black}$\circ$}} at  1.48149  0.55630
\put {{\color{black}$\circ$}} at  1.12494  0.60206
\put {{\color{black}$\circ$}} at  1.31876  0.64345
\put {{\color{black}$\circ$}} at  1.25181  0.64345
\put {{\color{black}$\circ$}} at  1.18709  0.69019
\put {{\color{black}$\circ$}} at  1.29243  1.63347
\put {{\color{black}$\circ$}} at  1.32790  1.65321
\put {{\color{black}$\circ$}} at  1.20761  0.71600
\put {{\color{black}$\circ$}} at  1.36653  0.77815
\put {{\color{black}$\circ$}} at  1.25181  0.81291
\put {{\color{black}$\circ$}} at  1.38722  0.85733
\put {{\color{black}$\circ$}} at  1.25181  0.88649
\put {{\color{black}$\circ$}} at  1.22185  0.91907
\put {{\color{black}$\circ$}} at  1.37675  0.92427
\put {{\color{black}$\circ$}} at  1.18046  1.94448
\put {{\color{black}$\circ$}} at  1.20066  0.98677

\setcoordinatesystem units <44.06mm,21.65mm> point at -1.3 -0.65
\setplotarea x from 1 to 2 , y from -1.0 to 1.0
\put {\includegraphics[width=4.406cm, height=2.165cm, bb=0 120 567 361]{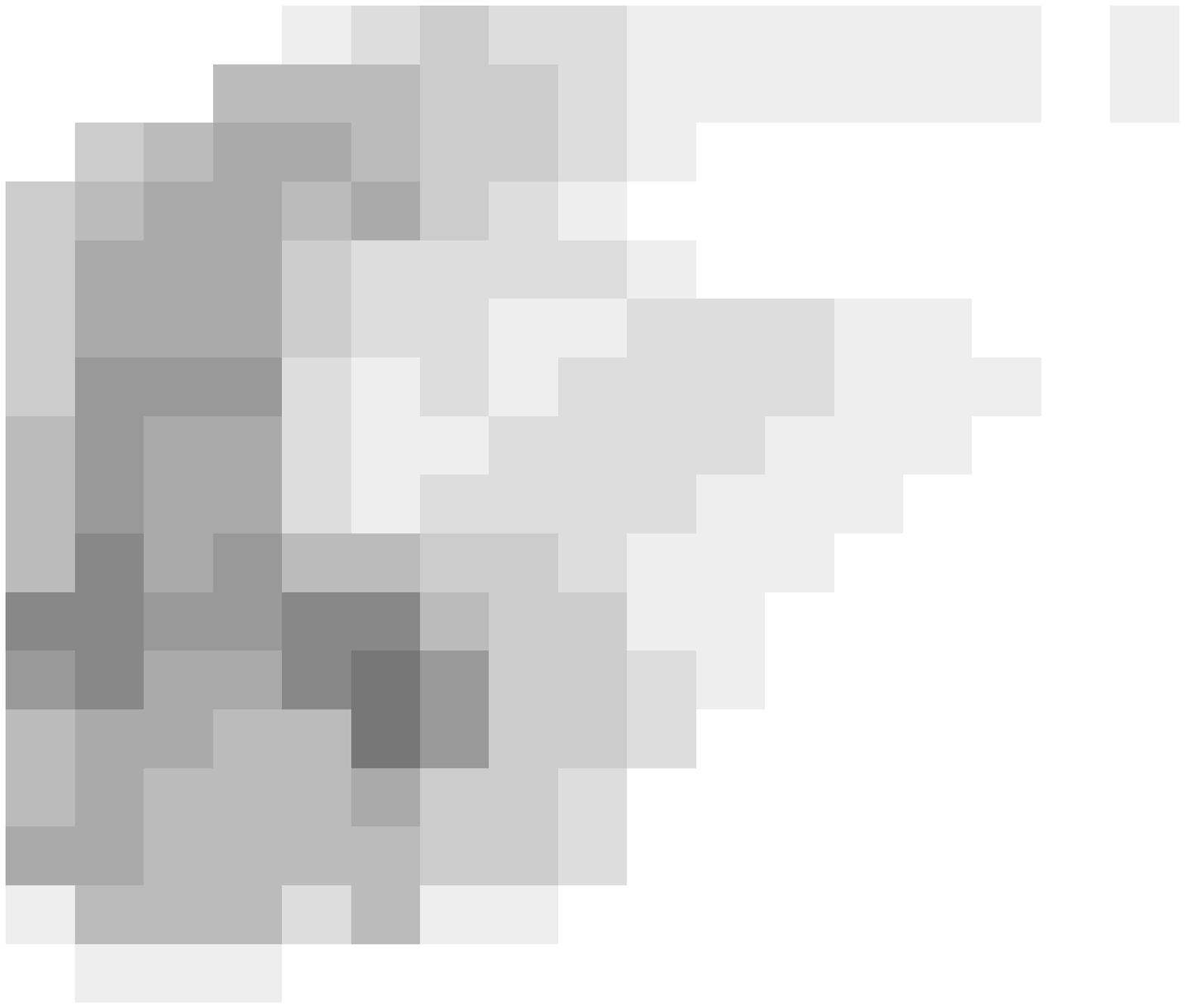}} at 1.5 0.0
\axis left label {}
ticks in long numbered from -1 to 1 by 1
      short unlabeled from -1 to 1 by 0.5 /
\axis right label {}
ticks in long unlabeled from -1 to 1 by 0.5
      short unlabeled from -1 to 1 by 0.5 /
\axis bottom label {}
ticks in long unlabeled from 1 to 2 by 0.2
      short unlabeled from 1 to 2 by 0.1 /
\axis top label {}
ticks in long unlabeled from 1.0 to 2 by 0.2
      short unlabeled from 1 to 2 by 0.1 /
\put {2.0} at 1.0 -1.15
\put {1.8} at 1.2 -1.15
\put {1.6} at 1.4 -1.15
\put {1.4} at 1.6 -1.15
\put {1.2} at 1.8 -1.15
\put {1.0} at 2.0 -1.15
\put {log[T$_{\rm bol}$/K]} at 1.5 -1.35
\put {\begin{sideways} log[M$_{\rm env}$/M$_\odot$] \end{sideways}} at 0.87 0.0
\put {{\color{black}$\circ$}} at 1.21467 -0.79588 
\put {{\color{black}$\circ$}} at 1.13668 -0.88605 
\put {{\color{black}$\circ$}} at 1.17393 -0.61978 
\put {{\color{black}$\circ$}} at 1.26761 -0.15490 
\put {{\color{black}$\circ$}} at 1.13668 -0.36653 
\put {{\color{black}$\circ$}} at 1.36653  0.49136 
\put {{\color{black}$\circ$}} at 1.37675  0.76342 
\put {{\color{black}$\circ$}} at 1.52288  0.07918 
\put {{\color{black}$\circ$}} at 1.38722  0.66275 
\put {{\color{black}$\circ$}} at 1.10791  0.00000 
\put {{\color{black}$\circ$}} at 1.34679 -0.14266 
\put {{\color{black}$\circ$}} at 1.48149 -0.09691 
\put {{\color{black}$\circ$}} at 1.12494 -0.44369 
\put {{\color{black}$\circ$}} at 1.31876 -0.14266 
\put {{\color{black}$\circ$}} at 1.25181  0.04139 
\put {{\color{black}$\circ$}} at 1.18709  0.00000 
\put {{\color{black}$\circ$}} at 1.29243  0.17609 
\put {{\color{black}$\circ$}} at 1.32790  0.55630 
\put {{\color{black}$\circ$}} at 1.20761  0.17609 
\put {{\color{black}$\circ$}} at 1.36653  0.00000 
\put {{\color{black}$\circ$}} at 1.25181 -0.28399 
\put {{\color{black}$\circ$}} at 1.38722  0.07918 
\put {{\color{black}$\circ$}} at 1.25181 -0.55284 
\put {{\color{black}$\circ$}} at 1.22185  0.04139 
\put {{\color{black}$\circ$}} at 1.37675  0.07918 
\put {{\color{black}$\circ$}} at 1.18046  0.47712 
\put {{\color{black}$\circ$}} at 1.20066 -0.34678 

\setcoordinatesystem units <15.20mm,21.65mm> point at -11. -0.65
\setplotarea x from -0.5 to 2.4 , y from -1.0 to 1.0
\put {\includegraphics[width=1.52cm, height=2.165cm, bb=193 120 395 360]{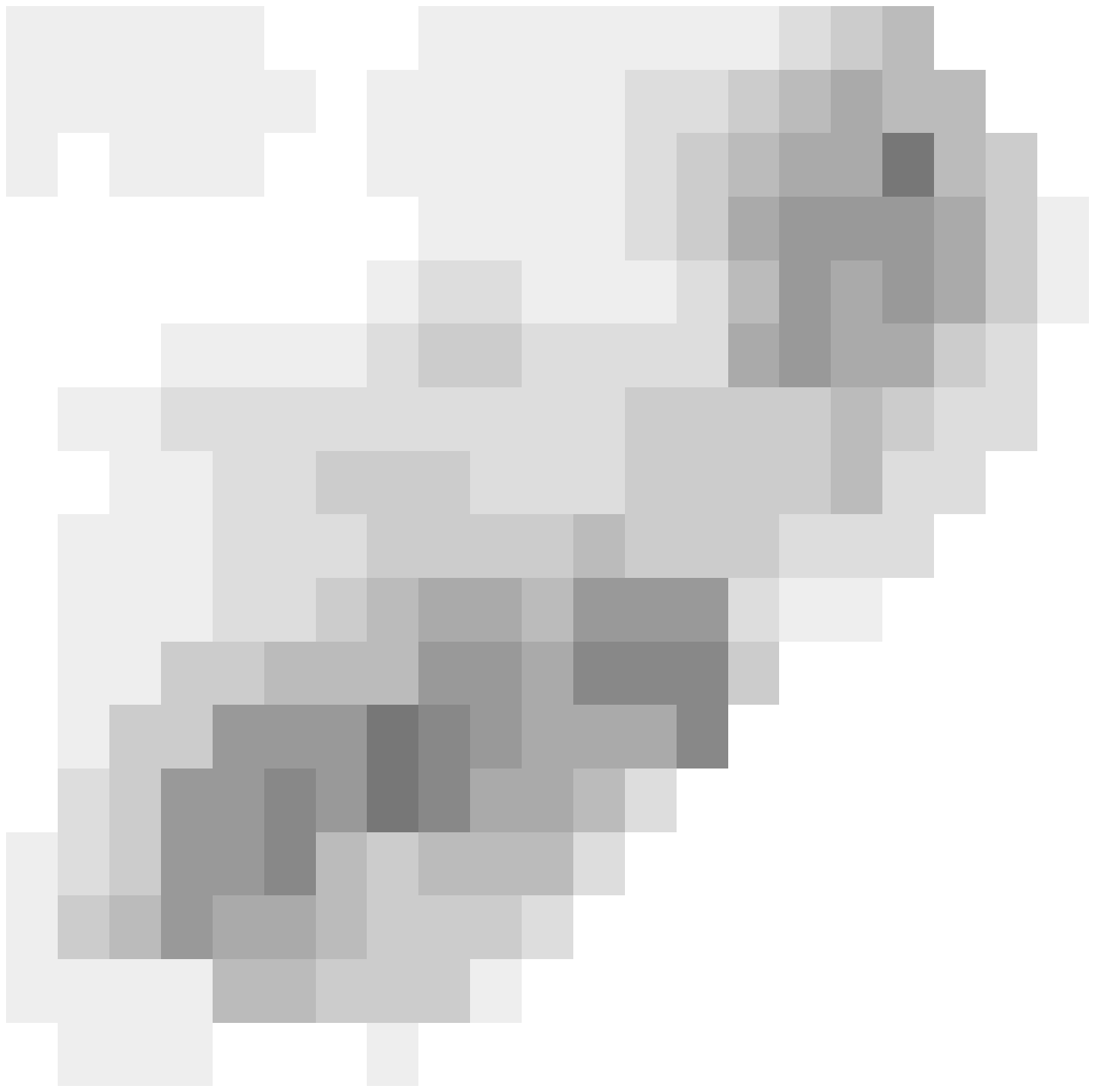}} at 0.95 0.0
\axis left label {}
ticks in long numbered from -1 to 1 by 1
      short unlabeled from -1 to 1 by 0.5 /
\axis right label {}
ticks in long unlabeled from -1 to 1 by 0.5
      short unlabeled from -1 to 1 by 0.5 /
\axis bottom label {}
ticks in long unlabeled from -0.0 to 2 by 1
      short unlabeled from -0.5 to 2. by 0.5 /
\axis top label {}
ticks in long unlabeled from -0.0 to 2 by 1
      short unlabeled from -0.5 to 2. by 0.5 /
\put {0} at 0 -1.15
\put {1} at 1 -1.15
\put {2} at 2 -1.15
\put {log[L$_{\rm bol}$/L$_\odot$]} at 0.95 -1.35
\put {\begin{sideways} log[M$_{\rm env}$/M$_\odot$] \end{sideways}} at -0.9 0.0
\put {{\color{black}$\circ$}} at -0.10790 -0.79588 
\put {{\color{black}$\circ$}} at -0.07058 -0.88605 
\put {{\color{black}$\circ$}} at  0.07918 -0.61978 
\put {{\color{black}$\circ$}} at  1.00000 -0.15490 
\put {{\color{black}$\circ$}} at  1.17609 -0.36653 
\put {{\color{black}$\circ$}} at  1.23045  0.49136 
\put {{\color{black}$\circ$}} at  1.25527  0.76342 
\put {{\color{black}$\circ$}} at  0.44715  0.07918 
\put {{\color{black}$\circ$}} at  1.32222  0.66275 
\put {{\color{black}$\circ$}} at  1.36173  0.00000 
\put {{\color{black}$\circ$}} at  0.49136 -0.14266 
\put {{\color{black}$\circ$}} at  0.55630 -0.09691 
\put {{\color{black}$\circ$}} at  0.60206 -0.44369 
\put {{\color{black}$\circ$}} at  0.64345 -0.14266 
\put {{\color{black}$\circ$}} at  0.64345  0.04139 
\put {{\color{black}$\circ$}} at  0.69019  0.00000 
\put {{\color{black}$\circ$}} at  1.63347  0.17609 
\put {{\color{black}$\circ$}} at  1.65321  0.55630 
\put {{\color{black}$\circ$}} at  0.71600  0.17609 
\put {{\color{black}$\circ$}} at  0.77815  0.00000 
\put {{\color{black}$\circ$}} at  0.81291 -0.28399 
\put {{\color{black}$\circ$}} at  0.85733  0.07918 
\put {{\color{black}$\circ$}} at  0.88649 -0.55284 
\put {{\color{black}$\circ$}} at  0.91907  0.04139 
\put {{\color{black}$\circ$}} at  0.92427  0.07918 
\put {{\color{black}$\circ$}} at  1.94448  0.47712 
\put {{\color{black}$\circ$}} at  0.98677 -0.34678 
\endpicture

\beginpicture
\setcoordinatesystem units <44.06mm,15.20mm> point at 0 0
\setplotarea x from 1 to 2 , y from -0.5 to 2.4
\put {\includegraphics[width=4.406cm, height=1.52cm, bb=0 157 566 323]{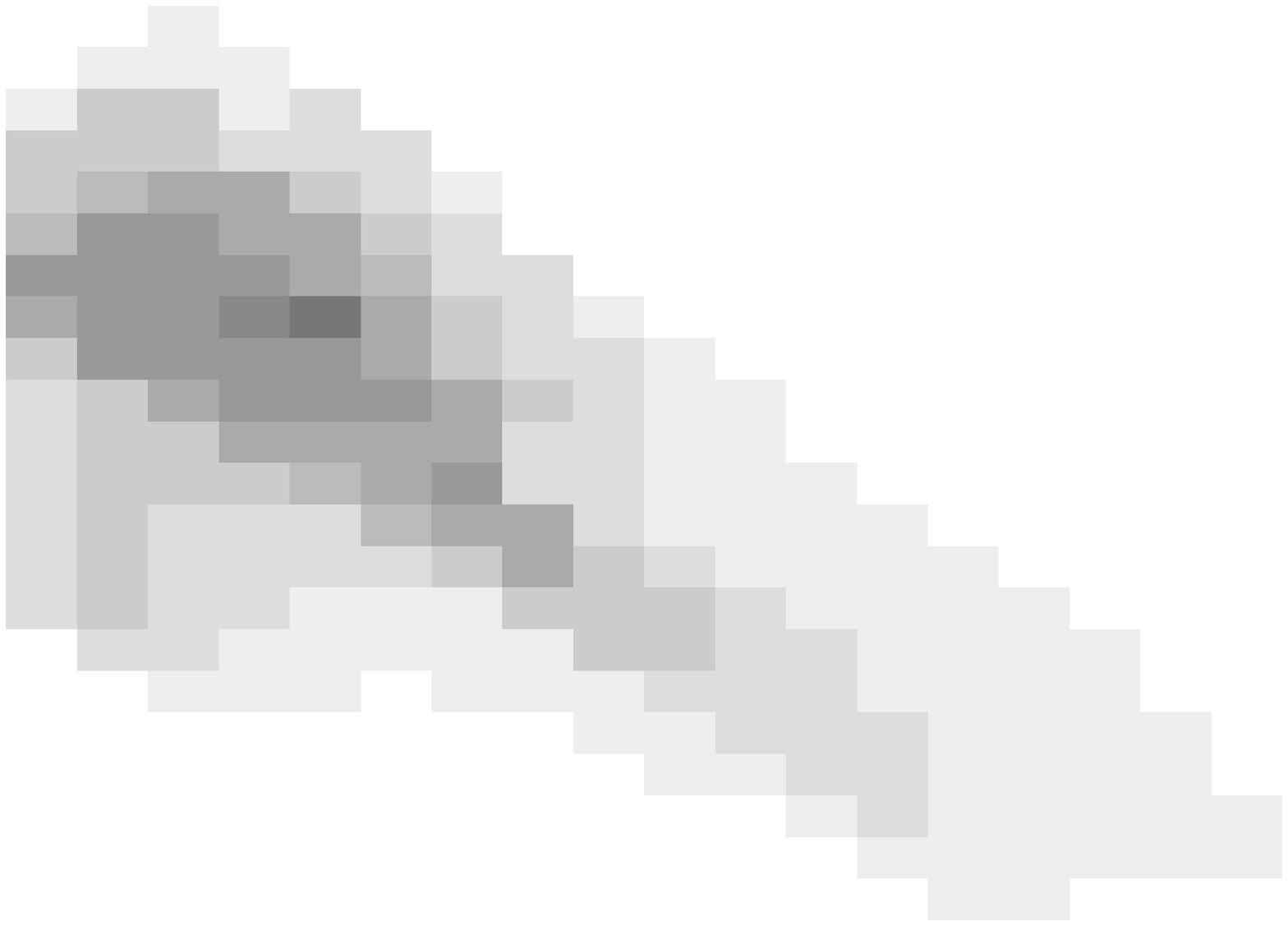}} at 1.5 0.95
\axis left label {}
ticks in long numbered from -0.0 to 2 by 1
      short unlabeled from -0.5 to 2. by 0.5 /
\axis right label {}
ticks in long unlabeled from -0.0 to 2 by 1
      short unlabeled from -0.5 to 2. by 0.5 /
\axis bottom label {}
ticks in long unlabeled from 1 to 2 by 0.2
      short unlabeled from 1 to 2 by 0.1 /
\axis top label {}
ticks in long unlabeled from 1.0 to 2 by 0.2
      short unlabeled from 1 to 2 by 0.1 /
\put {2.0} at 1.0 -0.7
\put {1.8} at 1.2 -0.7
\put {1.6} at 1.4 -0.7
\put {1.4} at 1.6 -0.7
\put {1.2} at 1.8 -0.7
\put {1.0} at 2.0 -0.7
\put {log[T$_{\rm bol}$/K]} at 1.5 -1.
\put {\begin{sideways} log[L$_{\rm bol}$/L$_\odot$] \end{sideways}} at 0.87 0.95
\put {{\color{black}$\circ$}} at  1.21467 -0.10790
\put {{\color{black}$\circ$}} at  1.13668 -0.07058
\put {{\color{black}$\circ$}} at  1.17393  0.07918
\put {{\color{black}$\circ$}} at  1.26761  1.00000
\put {{\color{black}$\circ$}} at  1.13668  1.17609
\put {{\color{black}$\circ$}} at  1.36653  1.23045
\put {{\color{black}$\circ$}} at  1.37675  1.25527
\put {{\color{black}$\circ$}} at  1.52288  0.44715
\put {{\color{black}$\circ$}} at  1.38722  1.32222
\put {{\color{black}$\circ$}} at  1.10791  1.36173
\put {{\color{black}$\circ$}} at  1.34679  0.49136
\put {{\color{black}$\circ$}} at  1.48149  0.55630
\put {{\color{black}$\circ$}} at  1.12494  0.60206
\put {{\color{black}$\circ$}} at  1.31876  0.64345
\put {{\color{black}$\circ$}} at  1.25181  0.64345
\put {{\color{black}$\circ$}} at  1.18709  0.69019
\put {{\color{black}$\circ$}} at  1.29243  1.63347
\put {{\color{black}$\circ$}} at  1.32790  1.65321
\put {{\color{black}$\circ$}} at  1.20761  0.71600
\put {{\color{black}$\circ$}} at  1.36653  0.77815
\put {{\color{black}$\circ$}} at  1.25181  0.81291
\put {{\color{black}$\circ$}} at  1.38722  0.85733
\put {{\color{black}$\circ$}} at  1.25181  0.88649
\put {{\color{black}$\circ$}} at  1.22185  0.91907
\put {{\color{black}$\circ$}} at  1.37675  0.92427
\put {{\color{black}$\circ$}} at  1.18046  1.94448
\put {{\color{black}$\circ$}} at  1.20066  0.98677

\setcoordinatesystem units <44.06mm,21.65mm> point at -1.3 -0.65
\setplotarea x from 1 to 2 , y from -1.0 to 1.0
\put {\includegraphics[width=4.406cm, height=2.165cm, bb=0 120 567 361]{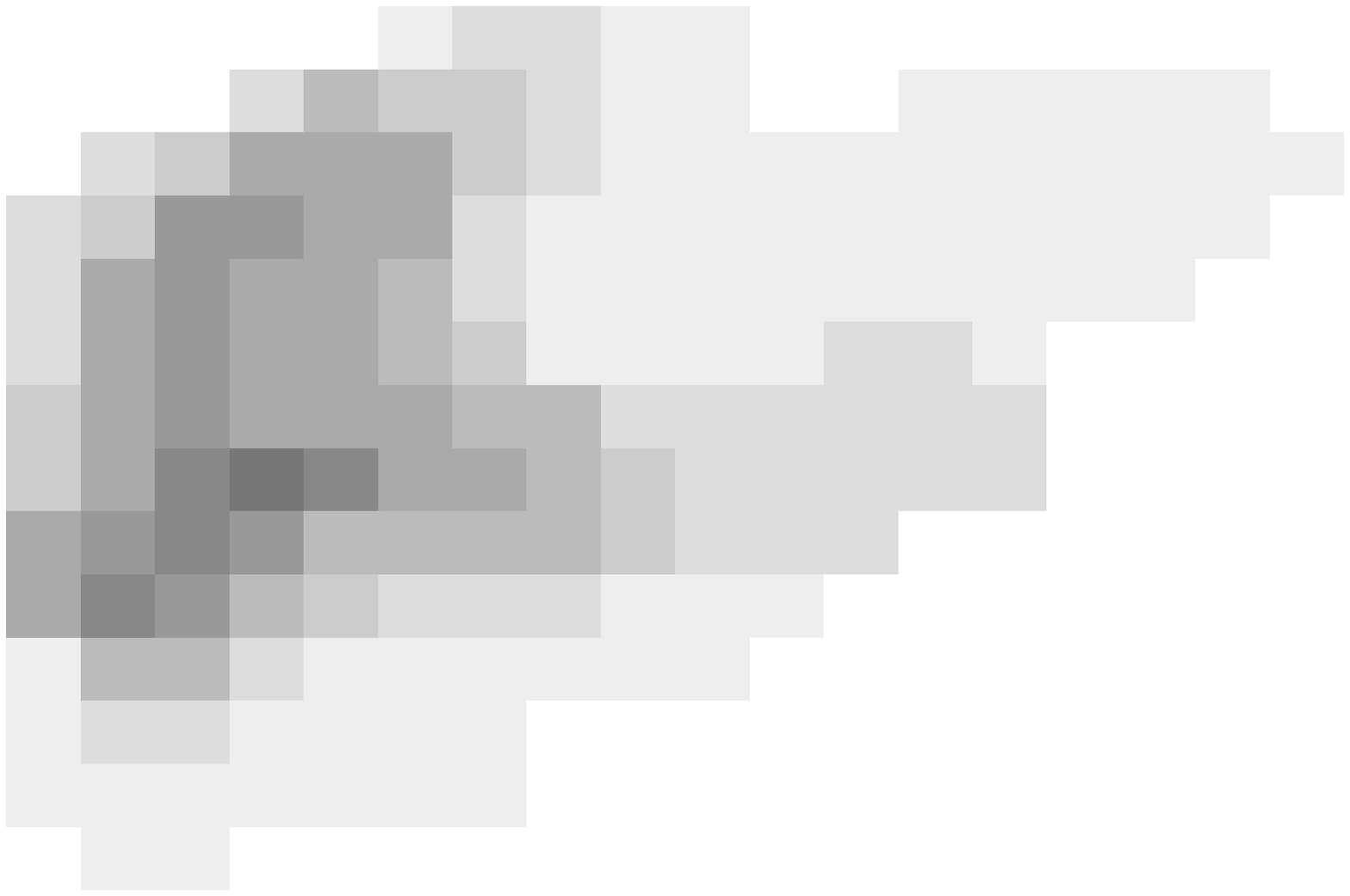}} at 1.5 0.0
\axis left label {}
ticks in long numbered from -1 to 1 by 1
      short unlabeled from -1 to 1 by 0.5 /
\axis right label {}
ticks in long unlabeled from -1 to 1 by 0.5
      short unlabeled from -1 to 1 by 0.5 /
\axis bottom label {}
ticks in long unlabeled from 1 to 2 by 0.2
      short unlabeled from 1 to 2 by 0.1 /
\axis top label {}
ticks in long unlabeled from 1.0 to 2 by 0.2
      short unlabeled from 1 to 2 by 0.1 /
\put {2.0} at 1.0 -1.15
\put {1.8} at 1.2 -1.15
\put {1.6} at 1.4 -1.15
\put {1.4} at 1.6 -1.15
\put {1.2} at 1.8 -1.15
\put {1.0} at 2.0 -1.15
\put {log[T$_{\rm bol}$/K]} at 1.5 -1.35
\put {\begin{sideways} log[M$_{\rm env}$/M$_\odot$] \end{sideways}} at 0.87 0.0
\put {{\color{black}$\circ$}} at 1.21467 -0.79588 
\put {{\color{black}$\circ$}} at 1.13668 -0.88605 
\put {{\color{black}$\circ$}} at 1.17393 -0.61978 
\put {{\color{black}$\circ$}} at 1.26761 -0.15490 
\put {{\color{black}$\circ$}} at 1.13668 -0.36653 
\put {{\color{black}$\circ$}} at 1.36653  0.49136 
\put {{\color{black}$\circ$}} at 1.37675  0.76342 
\put {{\color{black}$\circ$}} at 1.52288  0.07918 
\put {{\color{black}$\circ$}} at 1.38722  0.66275 
\put {{\color{black}$\circ$}} at 1.10791  0.00000 
\put {{\color{black}$\circ$}} at 1.34679 -0.14266 
\put {{\color{black}$\circ$}} at 1.48149 -0.09691 
\put {{\color{black}$\circ$}} at 1.12494 -0.44369 
\put {{\color{black}$\circ$}} at 1.31876 -0.14266 
\put {{\color{black}$\circ$}} at 1.25181  0.04139 
\put {{\color{black}$\circ$}} at 1.18709  0.00000 
\put {{\color{black}$\circ$}} at 1.29243  0.17609 
\put {{\color{black}$\circ$}} at 1.32790  0.55630 
\put {{\color{black}$\circ$}} at 1.20761  0.17609 
\put {{\color{black}$\circ$}} at 1.36653  0.00000 
\put {{\color{black}$\circ$}} at 1.25181 -0.28399 
\put {{\color{black}$\circ$}} at 1.38722  0.07918 
\put {{\color{black}$\circ$}} at 1.25181 -0.55284 
\put {{\color{black}$\circ$}} at 1.22185  0.04139 
\put {{\color{black}$\circ$}} at 1.37675  0.07918 
\put {{\color{black}$\circ$}} at 1.18046  0.47712 
\put {{\color{black}$\circ$}} at 1.20066 -0.34678 

\setcoordinatesystem units <15.20mm,21.65mm> point at -11. -0.65
\setplotarea x from -0.5 to 2.4 , y from -1.0 to 1.0
\put {\includegraphics[width=1.52cm, height=2.165cm, bb=193 120 395 360]{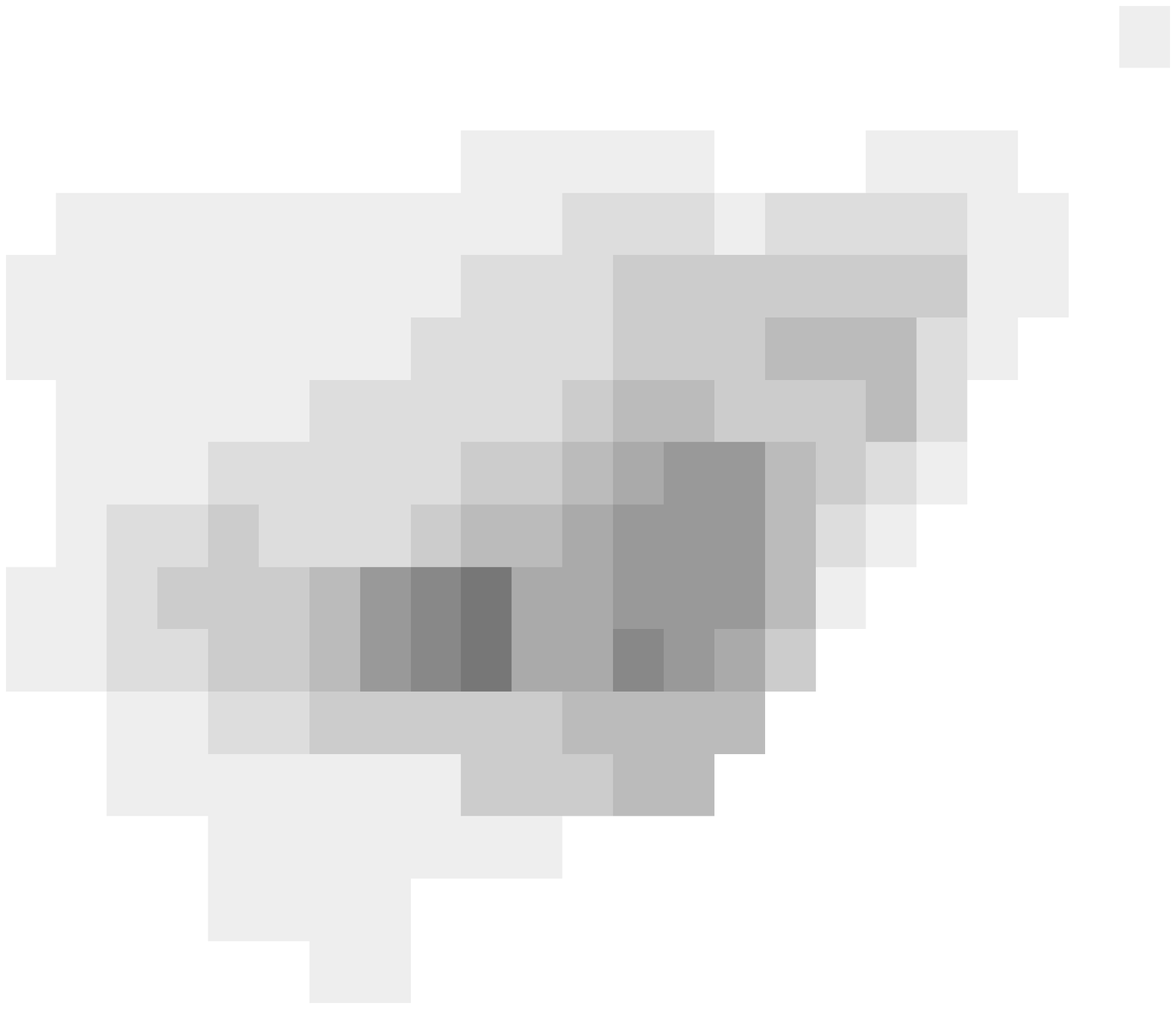}} at 0.95 0.0
\axis left label {}
ticks in long numbered from -1 to 1 by 1
      short unlabeled from -1 to 1 by 0.5 /
\axis right label {}
ticks in long unlabeled from -1 to 1 by 0.5
      short unlabeled from -1 to 1 by 0.5 /
\axis bottom label {}
ticks in long unlabeled from -0.0 to 2 by 1
      short unlabeled from -0.5 to 2. by 0.5 /
\axis top label {}
ticks in long unlabeled from -0.0 to 2 by 1
      short unlabeled from -0.5 to 2. by 0.5 /
\put {0} at 0 -1.15
\put {1} at 1 -1.15
\put {2} at 2 -1.15
\put {log[L$_{\rm bol}$/L$_\odot$]} at 0.95 -1.35
\put {\begin{sideways} log[M$_{\rm env}$/M$_\odot$] \end{sideways}} at -0.9 0.0
\put {{\color{black}$\circ$}} at -0.10790 -0.79588 
\put {{\color{black}$\circ$}} at -0.07058 -0.88605 
\put {{\color{black}$\circ$}} at  0.07918 -0.61978 
\put {{\color{black}$\circ$}} at  1.00000 -0.15490 
\put {{\color{black}$\circ$}} at  1.17609 -0.36653 
\put {{\color{black}$\circ$}} at  1.23045  0.49136 
\put {{\color{black}$\circ$}} at  1.25527  0.76342 
\put {{\color{black}$\circ$}} at  0.44715  0.07918 
\put {{\color{black}$\circ$}} at  1.32222  0.66275 
\put {{\color{black}$\circ$}} at  1.36173  0.00000 
\put {{\color{black}$\circ$}} at  0.49136 -0.14266 
\put {{\color{black}$\circ$}} at  0.55630 -0.09691 
\put {{\color{black}$\circ$}} at  0.60206 -0.44369 
\put {{\color{black}$\circ$}} at  0.64345 -0.14266 
\put {{\color{black}$\circ$}} at  0.64345  0.04139 
\put {{\color{black}$\circ$}} at  0.69019  0.00000 
\put {{\color{black}$\circ$}} at  1.63347  0.17609 
\put {{\color{black}$\circ$}} at  1.65321  0.55630 
\put {{\color{black}$\circ$}} at  0.71600  0.17609 
\put {{\color{black}$\circ$}} at  0.77815  0.00000 
\put {{\color{black}$\circ$}} at  0.81291 -0.28399 
\put {{\color{black}$\circ$}} at  0.85733  0.07918 
\put {{\color{black}$\circ$}} at  0.88649 -0.55284 
\put {{\color{black}$\circ$}} at  0.91907  0.04139 
\put {{\color{black}$\circ$}} at  0.92427  0.07918 
\put {{\color{black}$\circ$}} at  1.94448  0.47712 
\put {{\color{black}$\circ$}} at  0.98677 -0.34678 
\endpicture

\beginpicture
\setcoordinatesystem units <44.06mm,15.20mm> point at 0 0
\setplotarea x from 1 to 2 , y from -0.5 to 2.4
\put {\includegraphics[width=4.406cm, height=1.52cm, bb=0 157 566 323]{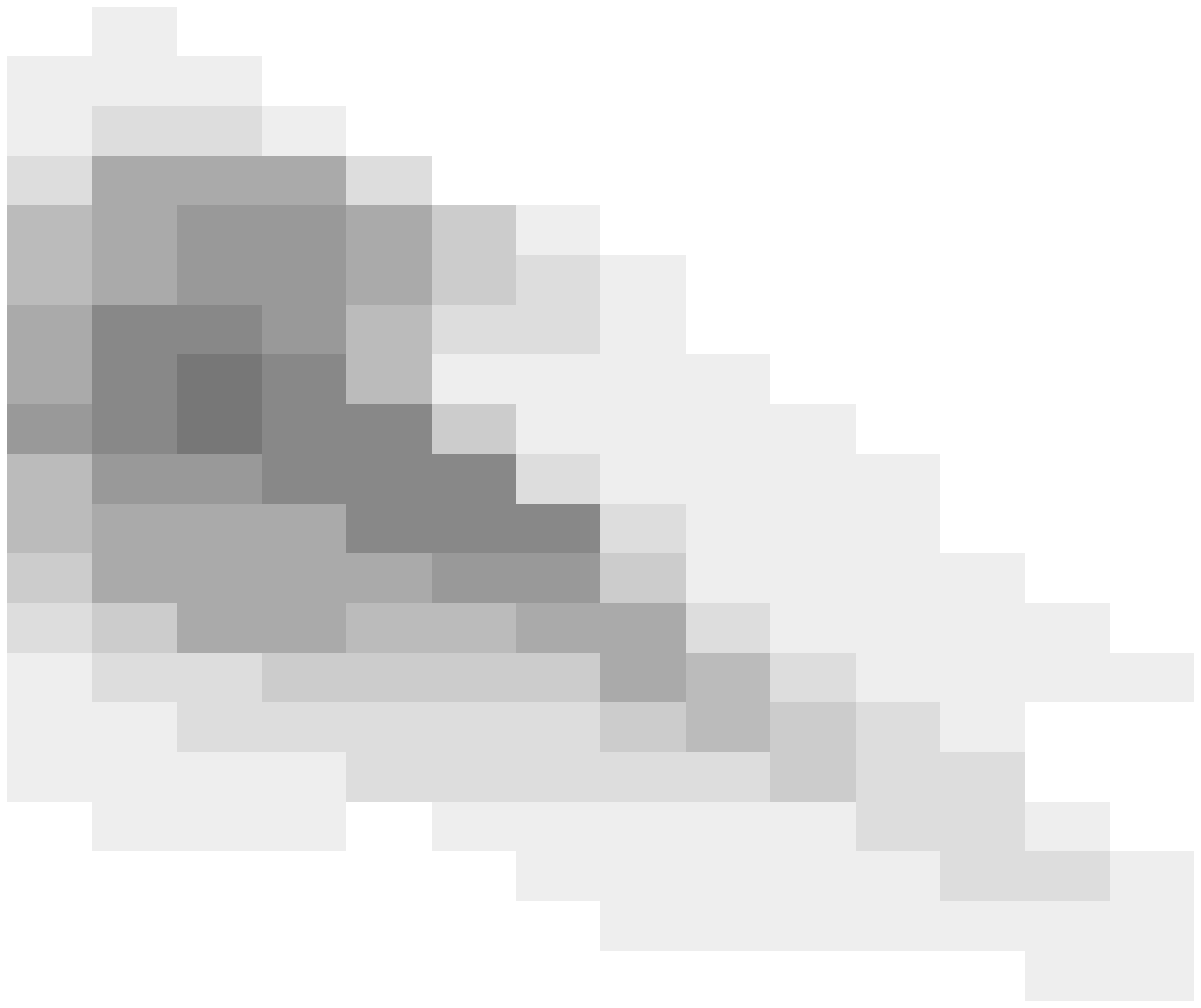}} at 1.5 0.95
\axis left label {}
ticks in long numbered from -0.0 to 2 by 1
      short unlabeled from -0.5 to 2. by 0.5 /
\axis right label {}
ticks in long unlabeled from -0.0 to 2 by 1
      short unlabeled from -0.5 to 2. by 0.5 /
\axis bottom label {}
ticks in long unlabeled from 1 to 2 by 0.2
      short unlabeled from 1 to 2 by 0.1 /
\axis top label {}
ticks in long unlabeled from 1.0 to 2 by 0.2
      short unlabeled from 1 to 2 by 0.1 /
\put {2.0} at 1.0 -0.7
\put {1.8} at 1.2 -0.7
\put {1.6} at 1.4 -0.7
\put {1.4} at 1.6 -0.7
\put {1.2} at 1.8 -0.7
\put {1.0} at 2.0 -0.7
\put {log[T$_{\rm bol}$/K]} at 1.5 -1.
\put {\begin{sideways} log[L$_{\rm bol}$/L$_\odot$] \end{sideways}} at 0.87 0.95
\put {{\color{black}$\circ$}} at  1.21467 -0.10790
\put {{\color{black}$\circ$}} at  1.13668 -0.07058
\put {{\color{black}$\circ$}} at  1.17393  0.07918
\put {{\color{black}$\circ$}} at  1.26761  1.00000
\put {{\color{black}$\circ$}} at  1.13668  1.17609
\put {{\color{black}$\circ$}} at  1.36653  1.23045
\put {{\color{black}$\circ$}} at  1.37675  1.25527
\put {{\color{black}$\circ$}} at  1.52288  0.44715
\put {{\color{black}$\circ$}} at  1.38722  1.32222
\put {{\color{black}$\circ$}} at  1.10791  1.36173
\put {{\color{black}$\circ$}} at  1.34679  0.49136
\put {{\color{black}$\circ$}} at  1.48149  0.55630
\put {{\color{black}$\circ$}} at  1.12494  0.60206
\put {{\color{black}$\circ$}} at  1.31876  0.64345
\put {{\color{black}$\circ$}} at  1.25181  0.64345
\put {{\color{black}$\circ$}} at  1.18709  0.69019
\put {{\color{black}$\circ$}} at  1.29243  1.63347
\put {{\color{black}$\circ$}} at  1.32790  1.65321
\put {{\color{black}$\circ$}} at  1.20761  0.71600
\put {{\color{black}$\circ$}} at  1.36653  0.77815
\put {{\color{black}$\circ$}} at  1.25181  0.81291
\put {{\color{black}$\circ$}} at  1.38722  0.85733
\put {{\color{black}$\circ$}} at  1.25181  0.88649
\put {{\color{black}$\circ$}} at  1.22185  0.91907
\put {{\color{black}$\circ$}} at  1.37675  0.92427
\put {{\color{black}$\circ$}} at  1.18046  1.94448
\put {{\color{black}$\circ$}} at  1.20066  0.98677

\setcoordinatesystem units <44.06mm,21.65mm> point at -1.3 -0.65
\setplotarea x from 1 to 2 , y from -1.0 to 1.0
\put {\includegraphics[width=4.406cm, height=2.165cm, bb=0 120 567 361]{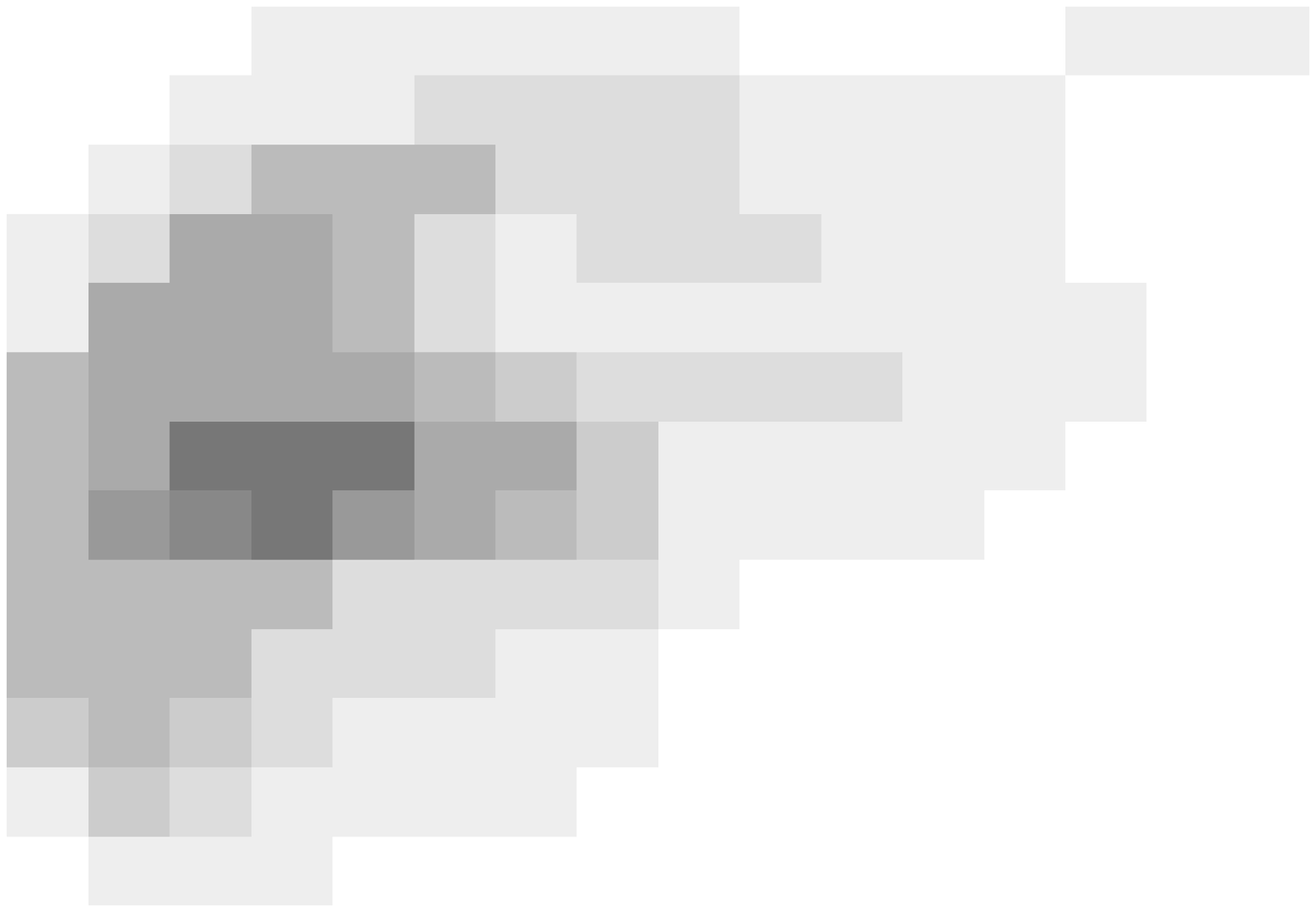}} at 1.5 0.0
\axis left label {}
ticks in long numbered from -1 to 1 by 1
      short unlabeled from -1 to 1 by 0.5 /
\axis right label {}
ticks in long unlabeled from -1 to 1 by 0.5
      short unlabeled from -1 to 1 by 0.5 /
\axis bottom label {}
ticks in long unlabeled from 1 to 2 by 0.2
      short unlabeled from 1 to 2 by 0.1 /
\axis top label {}
ticks in long unlabeled from 1.0 to 2 by 0.2
      short unlabeled from 1 to 2 by 0.1 /
\put {2.0} at 1.0 -1.15
\put {1.8} at 1.2 -1.15
\put {1.6} at 1.4 -1.15
\put {1.4} at 1.6 -1.15
\put {1.2} at 1.8 -1.15
\put {1.0} at 2.0 -1.15
\put {log[T$_{\rm bol}$/K]} at 1.5 -1.35
\put {\begin{sideways} log[M$_{\rm env}$/M$_\odot$] \end{sideways}} at 0.87 0.0
\put {{\color{black}$\circ$}} at 1.21467 -0.79588 
\put {{\color{black}$\circ$}} at 1.13668 -0.88605 
\put {{\color{black}$\circ$}} at 1.17393 -0.61978 
\put {{\color{black}$\circ$}} at 1.26761 -0.15490 
\put {{\color{black}$\circ$}} at 1.13668 -0.36653 
\put {{\color{black}$\circ$}} at 1.36653  0.49136 
\put {{\color{black}$\circ$}} at 1.37675  0.76342 
\put {{\color{black}$\circ$}} at 1.52288  0.07918 
\put {{\color{black}$\circ$}} at 1.38722  0.66275 
\put {{\color{black}$\circ$}} at 1.10791  0.00000 
\put {{\color{black}$\circ$}} at 1.34679 -0.14266 
\put {{\color{black}$\circ$}} at 1.48149 -0.09691 
\put {{\color{black}$\circ$}} at 1.12494 -0.44369 
\put {{\color{black}$\circ$}} at 1.31876 -0.14266 
\put {{\color{black}$\circ$}} at 1.25181  0.04139 
\put {{\color{black}$\circ$}} at 1.18709  0.00000 
\put {{\color{black}$\circ$}} at 1.29243  0.17609 
\put {{\color{black}$\circ$}} at 1.32790  0.55630 
\put {{\color{black}$\circ$}} at 1.20761  0.17609 
\put {{\color{black}$\circ$}} at 1.36653  0.00000 
\put {{\color{black}$\circ$}} at 1.25181 -0.28399 
\put {{\color{black}$\circ$}} at 1.38722  0.07918 
\put {{\color{black}$\circ$}} at 1.25181 -0.55284 
\put {{\color{black}$\circ$}} at 1.22185  0.04139 
\put {{\color{black}$\circ$}} at 1.37675  0.07918 
\put {{\color{black}$\circ$}} at 1.18046  0.47712 
\put {{\color{black}$\circ$}} at 1.20066 -0.34678 

\setcoordinatesystem units <15.20mm,21.65mm> point at -11. -0.65
\setplotarea x from -0.5 to 2.4 , y from -1.0 to 1.0
\put {\includegraphics[width=1.52cm, height=2.165cm, bb=193 120 395 360]{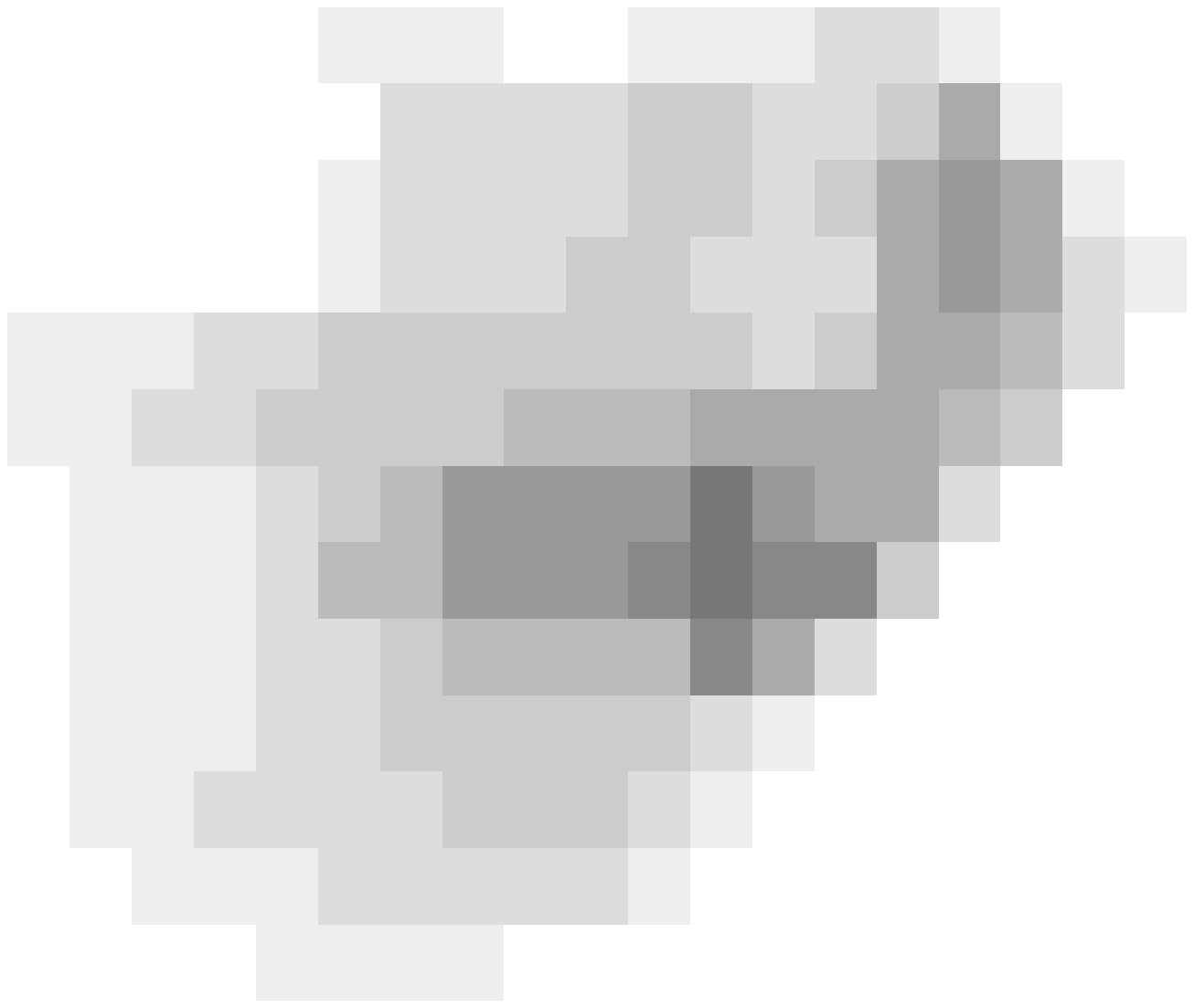}} at 0.95 0.0
\axis left label {}
ticks in long numbered from -1 to 1 by 1
      short unlabeled from -1 to 1 by 0.5 /
\axis right label {}
ticks in long unlabeled from -1 to 1 by 0.5
      short unlabeled from -1 to 1 by 0.5 /
\axis bottom label {}
ticks in long unlabeled from -0.0 to 2 by 1
      short unlabeled from -0.5 to 2. by 0.5 /
\axis top label {}
ticks in long unlabeled from -0.0 to 2 by 1
      short unlabeled from -0.5 to 2. by 0.5 /
\put {0} at 0 -1.15
\put {1} at 1 -1.15
\put {2} at 2 -1.15
\put {log[L$_{\rm bol}$/L$_\odot$]} at 0.95 -1.35
\put {\begin{sideways} log[M$_{\rm env}$/M$_\odot$] \end{sideways}} at -0.9 0.0
\put {{\color{black}$\circ$}} at -0.10790 -0.79588 
\put {{\color{black}$\circ$}} at -0.07058 -0.88605 
\put {{\color{black}$\circ$}} at  0.07918 -0.61978 
\put {{\color{black}$\circ$}} at  1.00000 -0.15490 
\put {{\color{black}$\circ$}} at  1.17609 -0.36653 
\put {{\color{black}$\circ$}} at  1.23045  0.49136 
\put {{\color{black}$\circ$}} at  1.25527  0.76342 
\put {{\color{black}$\circ$}} at  0.44715  0.07918 
\put {{\color{black}$\circ$}} at  1.32222  0.66275 
\put {{\color{black}$\circ$}} at  1.36173  0.00000 
\put {{\color{black}$\circ$}} at  0.49136 -0.14266 
\put {{\color{black}$\circ$}} at  0.55630 -0.09691 
\put {{\color{black}$\circ$}} at  0.60206 -0.44369 
\put {{\color{black}$\circ$}} at  0.64345 -0.14266 
\put {{\color{black}$\circ$}} at  0.64345  0.04139 
\put {{\color{black}$\circ$}} at  0.69019  0.00000 
\put {{\color{black}$\circ$}} at  1.63347  0.17609 
\put {{\color{black}$\circ$}} at  1.65321  0.55630 
\put {{\color{black}$\circ$}} at  0.71600  0.17609 
\put {{\color{black}$\circ$}} at  0.77815  0.00000 
\put {{\color{black}$\circ$}} at  0.81291 -0.28399 
\put {{\color{black}$\circ$}} at  0.85733  0.07918 
\put {{\color{black}$\circ$}} at  0.88649 -0.55284 
\put {{\color{black}$\circ$}} at  0.91907  0.04139 
\put {{\color{black}$\circ$}} at  0.92427  0.07918 
\put {{\color{black}$\circ$}} at  1.94448  0.47712 
\put {{\color{black}$\circ$}} at  0.98677 -0.34678 
\endpicture

\caption{\label{prob_all} Distribution of Class\,0 observations (circles) and
models (grey-scale) in the T$_{\rm bol}$-L$_{\rm bol}$ (left column), T$_{\rm
bol}$-M$_{\rm env}$ (middle column), and L$_{\rm bol}$-M$_{\rm env}$ (right
column) plane of the parameter space. The plots are shown for the models M05k8
(top row), M2k2 (middle row), and M6k2a (bottom row). Darker grey-values in the
model distribution corresponds to higher density of model points. As in
Fig.\,\ref{tracks} typical e-model parameters are used for the model
distributions (see Table\,\ref{best_agree}). Only model points with T$_{\rm
bol}$ smaller than 80\,K are shown in the plots since this is the limit of the
observational data.} 

\end{figure*}

\begin{table*}
\renewcommand{\tabcolsep}{4pt}
\centering
\caption{\label{best_agree} Parameter ranges needed in the e-model to best
match the observational data (see Appendix\,\ref{details_emod} for details on
the parameters). We list the values for the best fitting 5\,\% of the
parameter combinations. The last two columns list the median and average
duration of the Class\,0 phase of the individual stars in the models.} 
\begin{tabular}{lccccccccccccc}
Model & $T_{env}$ & $frac_{\rm env}$ & $M_{\rm extra}$ & $t_0$ & $\alpha$ &
$R_{\rm in}$ & $M_{\rm eff}$ & $p$ & $\kappa$ & $D_{\rm 3D}$ & $P_{\rm 3D}$ &
$t_{\rm Cl\,0}^{\rm med}$ & $t_{\rm Cl\,0}^{\rm ave}$ \\
\noalign{\smallskip}
 & [K] & [\%] & & [10$^3$\,yrs] & & [AU] & [\%] & & [cm$^2$g$^{-1}$] & & [\%] &
[10$^3$\,yrs] & [10$^3$\,yrs] \\
\noalign{\smallskip}
\hline
\noalign{\smallskip}
G2    &  15-19 & 81-89 & 0.6-2.4 & 35-75 & 1.4-3.2 & 30-80 & 44-47 & 1.6-1.9 & 3.0-5.2 & 0.30-0.37 & 37.2~-~5.93 & 41 & 44  \\    
M01k2 &  14-18 & 87-95 & 1.0-2.8 & 20-85 & 1.4-3.4 & 40-75 & 28-44 & 1.6-1.8 & 3.0-4.6 & 0.37-0.41 & 12.2~-~2.43 & 39 & 41  \\    
M05k4 &  17-20 & 89-94 & 1.2-2.0 & 35-75 & 1.6-3.2 & 45-85 & 41-44 & 1.5-1.7 & 3.0-4.4 & 0.29-0.37 & 40.3~-~4.72 & 31 & 36  \\    
M05k8 &  16-19 & 90-95 & 0.9-2.0 & 25-80 & 1.9-3.0 & 40-85 & 40-46 & 1.5-1.7 & 3.0-4.6 & 0.36-0.43 & 45.9~-~6.53 & 31 & 34  \\    
M2k2  &  15-20 & 86-92 & 1.0-2.4 & 35-85 & 1.6-3.2 & 35-80 & 35-43 & 1.5-1.7 & 3.4-5.2 & 0.29-0.36 & 73.1~-~19.6 & 32 & 39  \\    
M2k4  &  15-18 & 86-92 & 1.0-2.4 & 35-75 & 1.4-3.2 & 40-75 & 39-43 & 1.5-1.7 & 3.2-5.2 & 0.48-0.52 & 30.5~-~4.69 & 15 & 23  \\    
M2k8  &  13-18 & 80-90 & 1.0-2.2 & 30-75 & 1.2-3.6 & 35-75 & 32-41 & 1.5-1.7 & 3.4-5.4 & 0.27-0.34 & 67.4~-~17.0 & 45 & 56  \\    
M3k2  &  15-18 & 82-94 & 0.6-2.6 & 20-90 & 1.4-3.6 & 35-85 & 37-44 & 1.5-1.8 & 2.8-4.6 & 0.28-0.35 & 55.1~-~10.8 & 44 & 43  \\    
M3k4  &  12-14 & 80-92 & 0.6-1.6 & 25-80 & 1.4-3.4 & 30-70 & 32-44 & 1.5-1.8 & 3.2-5.2 & 0.33-0.40 & 5.99~-~0.53 & 87 & 105~ \\   
M6k2a &  16-20 & 85-95 & 1.0-2.4 & 20-85 & 1.4-3.4 & 45-85 & 35-43 & 1.6-1.8 & 2.8-4.6 & 0.32-0.36 & 46.5~-~20.4 & 26 & 32  \\    
M6k4a &  15-20 & 93-97 & 0.8-3.2 & 40-80 & 1.4-3.2 & 30-70 & 12-34 & 1.6-1.8 & 3.2-5.2 & 0.47-0.50 & 0.05~-~0.01 & 38 & 120~ \\   
M6k2b &  15-20 & 82-94 & 1.2-3.0 & 35-80 & 1.6-3.2 & 35-75 & 36-40 & 1.6-1.8 & 3.2-5.4 & 0.35-0.40 & 14.2~-~2.84 & 26 & 35  \\    
M6k2c &  15-20 & 86-96 & 1.0-3.0 & 35-75 & 1.4-3.2 & 35-75 & 36-42 & 1.6-1.9 & 3.4-4.8 & 0.38-0.43 & 5.30~-~0.62 & 29 & 51  \\    
M6k4c &  13-18 & 82-85 & 0.6-2.6 & 20-85 & 1.2-3.2 & 30-80 & 28-44 & 1.5-1.8 & 3.0-5.2 & 0.45-0.51 & 22.2~-~2.57 & 14 & 54  \\    
M10k2 &  16-22 & 92-96 & 1.0-3.6 & 25-70 & 1.2-3.0 & 35-80 & 10-42 & 1.6-1.9 & 3.2-5.4 & 0.60-0.62 & 0.003-~0.00 & 19 & 24  \\    
M10k8 &  14-17 & 93-98 & 0.8-3.2 & 30-85 & 0.8-3.8 & 35-80 & 16-38 & 1.5-1.7 & 3.4-5.2 & 0.43-0.49 & 2.60~-~0.23 & 44 & 124~ \\   
\end{tabular}
\end{table*}

\subsection{Distribution in T$_{\rm bol}$-L$_{\rm bol}$-M$_{\rm env}$}

In Fig.\,\ref{prob_all} we show how the model tracks (M05k8, M2k2, M6k2a, from
left to right) are distributed in the full T$_{\rm bol}$-L$_{\rm bol}$-M$_{\rm
env}$ parameter space. The grey-scale background of the individual panels gives
the probability to find one of the model stars at this particular position.
Darker regions indicate higher probabilities. We used typical e-model parameters
(see Table\,\ref{best_agree}) to determine these diagrams. For comparison the
observational datapoints are overplotted. Note that we plotted only model points
with T$_{\rm bol} < 80$\,K, as this is the limit of the observational data and
only these are used in the comparison with the observations. 

As evident in Fig.\,\ref{prob_all}, the model distributions cover roughly the
same area as the observations. Especially in the T$_{\rm bol}$-L$_{\rm bol}$ and
T$_{\rm bol}$-M$_{\rm env}$ plane the models are able to explain the peak in the
observed distribution. In the L$_{\rm bol}$-M$_{\rm env}$ plane, however, the
models fail to explain this peak (see right column in Fig.\,\ref{prob_all}). Our
models predict smaller envelope masses (by a factor of two) compared to parts of
the observations. Considering the cluster of observational points at about
M$_{\rm env} = 1$\,M$_\odot$, the discrepancy could be a selection effect in the
observations. In addition, observed envelope masses are uncertain by a factor of
two (Motte \& Andr\'e \cite{2001A&A...365..440M}). See also the discussion in
Sec.\,\ref{further}.

\subsection{Initial mass function}
\label{imf_comp}
 
We analyse the mass function of the final masses of model stars (IMF) in the
gt-models in order to compare it to that of the observational sample. The
protostellar mass functions of the gt-models show a decline for masses larger
than about 0.3\,M$_{\odot}$. The mean value of the slope of all considered
models is $\langle \Gamma \rangle$\,=\,$-0.84$ in the mass range
$-0.5$\,$<$\,$\log {\rm M}/{\rm M}_{\odot}$\,$<$\,1. This is less steep than the
Salpeter slope of $-1.35$ (Salpeter \cite{1955ApJ...121..161S}) but our results
are biased by the fact that binaries or multiple systems cannot be resolved. At
the low-mass end, the mass function is constrained by the SPH resolution limit
of $\sim$\,0.05\,M$_{\odot}$. Nevertheless, the mean value is in good agreement
with the estimated final stellar mass spectrum in the corresponding
observational sample ($\Gamma$\,=\,-0.9$\pm$0.2; Froebrich
\cite{2005ApJS..156..169F}). Recall, that for this sample also no binary
correction has been made.

We test if the slope of the IMF in the gt-models is related to the best
agreement ($P_{\rm 3D}$). However, no systematic dependence of $P_{\rm 3D}$ on
$\Gamma$ is present. In particular, a good agreement of $\Gamma$ between the
gt-model and the observations does not necessarily lead to a high $P_{\rm
3D}$-value. As discussed in the last paragraph (see also the middle panel of
Fig.\,\ref{tracks}), the differences in the position in the T$_{\rm
bol}$-L$_{\rm bol}$ diagram are partly due to the accretion history and not due
to the final mass. Hence, the test performed here is much more sensitive to the
distribution of individual accretion histories than to the slope of the IMF.

\subsection{Evolutionary model}
\label{disc_emod}
 
To constrain the free parameters in the e-model, we investigate here how the
chosen values influence the agreement $P_{\rm 3D}$. Table\,\ref{best_agree}
contains the range of values which provide the highest agreement for each
gt-model. In particular we show the values for the best fitting 5\,\% of the
e-models. It is more meaningful to provide these ranges instead of the
particular parameters for the single best fitting model. These are statistically
less significant, since not the whole parameter space is tested by our method.
We obtain the following results for all gt-models:

\begin{itemize}

\item The temperature at the outer envelope boundary is lower than the standard
value of 24\,K. Typically temperatures between 15 and 19\,K lead to a good
agreement. 

\item A wide range for inner envelope radii ($R_{\rm in}$) leads to high $P_{\rm
3D}$. Values from 35 to 80\,AU are typical. This wide range is in agreement with
the fact that a change in $R_{\rm in}$, while fixing all other parameters,
usually results in very small changes of $P_{\rm 3D}$. However, the range for
the best fitting models does not include the hitherto standard value of 30\,AU.

\item In most cases small amounts of extra mass in the envelope ($1.0 < M_{\rm
extra} < 2.4$) lead to high agreement.

\item The parameter $\alpha$ shows a wide range of possible values for a good
agreement (1.4\,$< \alpha <$\,3.2). A similar wide range is found for $t_0$
which ranges from 30 to 80\,$\cdot$\,10$^3$\,yrs. The larger values found for
$t_0$ suggest that the additional mass stays longer in the envelope to sustain
low bolometric temperatures. 

\item A range of values of $p$ is found to be able to best explain the data.
However, in the majority of the cases $1.55 < p < 1.80$ leads to the best
results. These values are in between the theoretical solutions for a free
falling envelope (1.5) and a singular isothermal sphere (2.0). The range found
here seems to be a compromise between younger and older objects. Since there are
probably more older sources in the sample, the best values are closer to
$p=1.5$. In principle an evolution from $p=2.0$ to $p=1.5$ would be expected.
Our findings do not contradict this. Our obtained values are in good agreement
with (sub)-mm observations of envelopes (Chandler \& Richer
\cite{2000ApJ...530..851C}; Motte \& Andr\'e \cite{2001A&A...365..440M}).

\item We found the mass fraction in the envelope to be in the range from 86 to
96\,\%. This would imply a disc mass of 4 to 14\,\% of the envelope mass, mostly
smaller than the standard value of 13\,\% used so far, but in agreement with
millimetre interferometric observations (e.g. Looney et al.
\cite{2003ApJ...592..255L}).

\item The amount of mass ejected into the jets is found to range from 36 to
44\,\%. This is a narrow range and somewhat contradictory to the observations
and jet launching models which predict that at maximum 30\,\% of the material
is ejected into the jets. We interpret the high amount of ejected material here
as a combination of two things. (1) Material really ejected into the jets. (2)
Material accreted onto the star, but the resulting luminosity escapes directly
through the cavity generated by the jets and is not observed. If the accretion
luminosity is radiated uniformly away, then the additional ejected material
would imply opening angles between 30 and 60 degrees for the cavities, not an
untypical value for many of the older Class\,0 sources (e.g. Padman et al.
\cite{1997IAUS..182..123P}). To conclusively prove this hypothesis one needs to
include the outflow luminosity for all objects into the comparison of models
and observations. 

\item The best-fitting opacity range (3.0 to 5.0\,cm$^2$g$^{-1}$) agrees very
well with the standard value of 4\,cm$^2$g$^{-1}$.

\end{itemize}

\subsection{Gravoturbulent models}
\begin{figure}
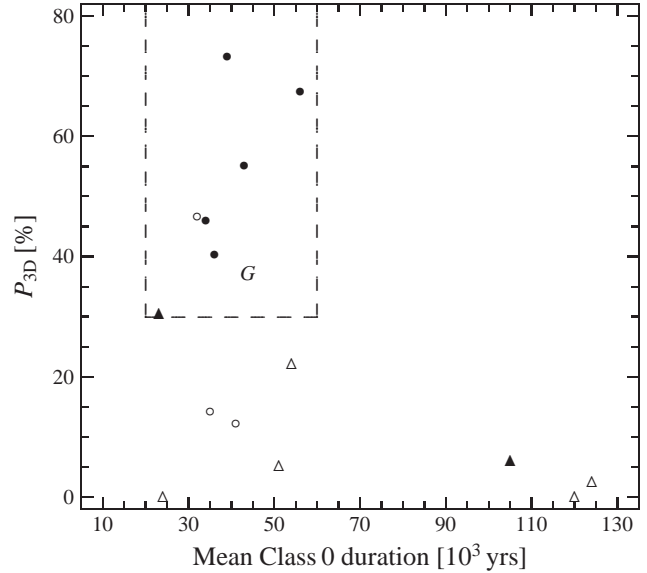

\beginpicture
\setcoordinatesystem units <0.57mm,0.8mm> point at 0 0
\setplotarea x from 5 to 135 , y from -2 to 82
\axis left label {}
ticks in long numbered from 0 to 80 by 20
      short unlabeled from 0 to 80 by 5 /
\axis right label {}
ticks in long unlabeled from 0 to 80 by 20
      short unlabeled from 0 to 80 by 5 /
\axis bottom label {}
ticks in long numbered from 10 to 130 by 20
      short unlabeled from 10 to 135 by 5 /
\axis top label {}
ticks in long unlabeled from 10 to 130 by 20
      short unlabeled from 10 to 135 by 5 /
\put {\large Mean Class\,0 duration [10$^3$\,yrs]} at 70 -10
\put {\large \begin{sideways}$P_{\rm 3D}$\,[\%]\end{sideways}} at -8 40
\put {$G$}                  at 44    37.2       
\put {$\circ$}              at 41    12.2       
\put {$\bullet$}            at 36    40.3       
\put {$\bullet$}            at 34    45.9       
\put {$\bullet$}            at 39    73.1       
\put {$\blacktriangle$}     at 23    30.5       
\put {$\bullet$}            at 56    67.4       
\put {$\bullet$}            at 43    55.1       
\put {$\blacktriangle$}     at 105   5.99       
\put {$\circ$}              at 32    46.5       
\put {$\vartriangle$}       at 120   0.05       
\put {$\circ$}              at 35    14.2       
\put {$\vartriangle$}       at 51    5.30       
\put {$\vartriangle$}       at 54    22.2       
\put {$\vartriangle$}       at 24    0.003      
\put {$\vartriangle$}       at 124   2.60       
\setdashes
\plot 20 30 20 82 /
\plot 60 30 60 82 /
\plot 20 30 60 30 /
\setsolid
\endpicture
\caption{\label{durationvsp} Mean duration of the Class\,0 phase of the model
stars vs. the $P_{\rm 3D}$ value for the gravoturbulent models. Circles mark
models that possess a median duration of the Class\,0 phase between 2 and
6\,$\cdot$\,10$^4$\,yrs and a ratio of median to mean duration of the Class\,0
phase of higher than 0.7. Triangles mark models were one of these conditions is
not fulfilled. Gravoturbulent models with Mach numbers in the range $0.5 \le 
\mathcal{M} \le 3.0$ are shown as filled symbols, models outside this range as
open symbols. The Gaussian collapse model is marked as `$G$'. The dashed lines
enclose the region where a high agreement of the models with the observations
was found.} 
\end{figure}
 
\label{discus_gtmod}

We now investigate if the initial conditions of the gt-models influence the
agreement with the observations and the best fitting e-model parameters. The
first point to note is that we find no significant correlations of any of the
obtained parameter values of the e-models with the Mach number or wavelength
that characterise the turbulence in the gt-model. 

There are, however, large differences in the quality of agreement between the
various gt-models and the observations. In the following we investigate the
cause of these differences. We determined for each set of accretion rates the
median and the mean duration of the Class\,0 phase of the model stars. The
median for all gt-models ranges from 14 to 87\,$\cdot$\,10$^3$yrs, while the
mean duration spans from 23 to 124\,$\cdot$\,10$^3$yrs. The much wider range for
the average duration is due to some model stars (outliers) that go through an
extremely long Class\,0 phase. These objects have a large, secondary accretion
peak which in many cases might not be physical but is caused by the lack of
feedback mechanisms. Due to the mass criterion used, such peaks shift the
transition from Class\,0 to 1 to later times. In Table\,\ref{best_agree} we list
the median and mean duration of the Class\,0 phase of all stars in the models in
the last two columns.

All gt-models with $P_{\rm 3D}$-values larger than 30\,\% are associated with a
mean duration of the Class\,0 phase between 2 and 6\,$\cdot$\,10$^4$yrs. Most
of them possess a ratio of median to mean duration of the Class\,0 phase
exceeding 0.7. This indicates that no or only a few outliers are present in the
set of accretion rates. When applying these two conditions to all gt-models, we
select nine of them (circles + $G$ in Fig.\,\ref{durationvsp}). Only M01k2
(12.2\,\%) and M6k2b (14.2\,\%) do not possess $P_{\rm 3D} > 30\,\%$. Hence
80\,\% of the gt-models with these properties lead to high agreement. There are
seven models that do not fulfil one of the criteria (triangles in
Fig.\,\ref{durationvsp}). Six of these possess $P_{\rm 3D}$-values smaller than
30\,\%. Hence 85\,\% of the gt-models that do not comply with both criteria lead
to low agreement. This shows that high $P_{\rm 3D}$-values are only obtained for
gt-models that generate few outliers and the `right' duration of the Class\,0
phase. 

The duration of the Class\,0 stage shows a linear correlation with the median of
the fit parameter $\tau$ of SK04. This parameter indicates the time it takes
until the peak accretion rate is reached. It is related to the local free-fall
time and, thus, can be used as a rough estimate of the average density of the
core at the onset of collapse (see SK04 for details). Given our best fitting
models and the corresponding Class\,0 lifetimes, we infer a range of
1-5\,$\cdot$10$^5$\,cm$^{-3}$ for the local density at the start of the
collapse. This is in good agreement with measured central densities of
pre-stellar cores (e.g. Bacmann et al. \cite{2002A&A...389L...6B}). Our results
thus suggest that stars in the observational sample may only form in pre-stellar
cloud cores that reach a central density of order of $10^5$\,cm$^{-3}$.
Obviously, the duration of the Class\,0 phase depends directly on the initial
local density. 

As can be seen in Table\,\ref{best_agree}, gt-models with Mach numbers $0.5 \le
{\mathcal M} \le 3$ best describe the observations (filled symbols in
Fig.\,\ref{durationvsp}). Six out of seven such models possess a high $P_{\rm
3D}$-value, while only one out of eight models with ${\mathcal M} \ge 6$ has a
high $P_{\rm 3D}$. The exceptions M3k4 (${\mathcal M} < 6$, $P_{\rm
3D}$=\,5.99\,\%) and M6k2a (${\mathcal M} \ge 6$, $P_{\rm 3D}$=\,46.5\,\%), as
well as the fact that similar gt-models (e.g. M6k2a/b/c) result in very
different $P_{\rm 3D}$-values are, however, understandable since protostellar
collapse is, in essence, a rather localised process, connected to the turbulent
cloud environment only via the available mass and angular momentum inflow rate.
It indicates that star formation is a highly stochastic process, influenced more
strongly by local properties than by the global initial conditions, a fact which
is also found, for example, by Bate et al. \cite{2003MNRAS.339..577B}.

The range of Mach numbers found as best explanation for the data reflects the
composition of the observational sample. It mainly contains sources from
Perseus, Orion, and Serpens (Froebrich \cite{2005ApJS..156..169F}). There
typical Mach numbers are in the range $1 \le {\mathcal M} \le 6$ (e.g. Castets
\& Langer \cite{1995A&A...294..835C}, Schmeja et al.
\cite{2005A&A...437..911S}). It is, however, interesting to note that the
majority of gt-models with ${\mathcal M} \ge 6$ are not able to explain the
observations, independent of the choice for the e-model parameters. This
indicates that the initial conditions do influence the observational properties
of Class\,0 protostars.

\subsection{Underluminous sources}

\label{underlumi}
 
We recall that we excluded 25\,\% of the observational data in our
determination of $P_{\rm 3D}$. Those excluded exhibit a much lower bolometric
luminosity than the other objects, for their given bolometric temperature and
envelope mass. They could be objects which experience either (a) quiescent
states of low accretion or (b) a different time dependence of the accretion
rate (Froebrich \cite{2005ApJS..156..169F}). 

A detailed investigation of the individual accretion rates from the gt-models
rules out the existence of quiet accretion phases as the dominant cause for the
low luminosity of these sources. On average, the individual model stars spend
less than 5\,\% of their time in a phase of suppressed accretion which we define
as being a factor of six lower than the average value. This implies that we
should observe only about 5\,\% of the sources in such a state, compared to
25\,\%. Hence, these objects appear to follow a different accretion history.

We investigate this in more detail by performing the procedure of determining
$P_{\rm 3D}$ using not only the `normal' sources, but also the Taurus-like
objects. We find that this reveals worse agreement with $P_{\rm 3D}$ values
typically halved. The agreement between models and observations is always lower
when we try to explain the Taurus like sources. This strengthens the proposal
that in the case of `normal' sources, turbulence governs the accretion process,
while other processes are responsible for the underluminous objects. Considering
their lower luminosity and hence lower accretion rates, we speculate that
ambipolar diffusion might be an important process governing accretion in these
sources. 

A more detailed analysis of a sub-sample of Taurus like objects would be
desirable since the time dependence of the mass accretion rate might vary from
region to region (Henriksen et al. \cite{1997A&A...323..549H}). Even though more
and more such objects are discovered (e.g. Young et al.
\cite{2004ApJS..154..396Y}), indicating that the known percentage of these
objects is clearly a lower limit, the current sample still suffers from too
small number statistics to perform a reliable statistical analysis.

\subsection{Further discussion}

\label{further}
 
As evident from Table\,\ref{best_agree}, the best agreement between observations
and models is rather low ($\approx$\,70\,\%). Here we will discuss three
possible causes for this:

\begin{enumerate}

\item[\bf 1. Our procedure:] There are many free parameters which, in principle,
could take values within a large range, as noted in Table\,\ref{params}. A
straightforward test of all possible combinations to solve this highly
non-linear minimisation problem of many variables is impossible. We thus applied
a Monte Carlo method to obtain the best fitting parameter combination (see
Appendix\,\ref{details_kstest}). The obtained probability $P_{\rm 3D}$ is a
lower limit since not all possible parameter combinations have been tested. Note
that only slightly smaller values for $D_{\rm 3D}$ can lead to probabilities
much closer to 100\,\%, especially for models that already possess a high
agreement.

When converting $D_{\rm 3D}$ to $P_{\rm 3D}$ we restricted the random selection
to model values with bolometric temperatures smaller than 80\,K. This is the
limit of our observational sample of Class\,0 sources. There are further
restrictions in the observational data. No objects outside a given range in
L$_{\rm bol}$ and M$_{\rm env}$. This is most likely an observational bias. We
refrain from applying the other observational limits for the random selection
process for the $P_{\rm 3D}$ determination because the observational limits are
not well defined/understood. If doing so, however, we would increase the
probabilities for the agreement in many cases significantly towards 100\,\%. 

\item[\bf 2. Observational data:] As stated above, the observations clearly
cover only a limited range of the T$_{\rm bol}$-L$_{\rm bol}$-M$_{\rm env}$
parameter space. All limits, except T$_{\rm bol} < 80$\,K, are observational
biases. Note that the observational sample consists of sources from several
star forming regions. It is possible that e.g. the sensitivity limit for the
bolometric luminosity is not the same in all regions, changing the observed
statistics of low luminosity/mass sources compared to the real one. 

As shown in Fig.\,\ref{tracks}, applying a fixed limit for T$_{\rm bol}$ as a
divide between Class\,0 and Class\,1 objects is not valid (see also Young \&
Evans \cite{2005ApJ...627..293Y}). Most of the solar mass stars undergo this
transition at temperatures close to 80\,K. However, depending on the individual
accretion history of each star, some quite significant deviations from this
value are evident. Furthermore, lower mass Class\,0/Class\,1 transition objects
tend to be warmer than the temperature limit applied here. Thus, the object
sample might suffer from misclassified objects. This also shows that there is a
difference in the definition for Class\,0 protostars based on observations and
models. Furthermore, the observational definition describes a different physical
state of the object depending on its accretion history and final mass. We
overcome most of these uncertainties, however, by applying the T$_{\rm bol} <
80$\,K restriction when determining $P_{\rm 3D}$.

\item[\bf 3. The models:] All our models are constructed on the most basic
principles (e.g. the gt-models neglect feedback mechanisms, the disc mass is a
constant fraction of the envelope mass, etc.). In addition, all model stars are
treated with the same set of e-model parameters. It is not clear if some of the
parameters might depend on the final mass of the star. Many parameters are also
kept constant in time. This might not be valid (e.g. the inner
envelope radius, the fraction of mass in the disc, and the powerlaw index of the
envelope density distribution could depend on the evolutionary state, final mass
of the star). It is in many cases, however, uncertain how these dependencies can
be parameterised. We thus kept these parameters constant so as not to introduce even
more free arbitrary parameters.

\end{enumerate}

Given these three points, we conclude that a very good agreement between the
models and observations should not be expected. It would rather be very worrying
if our method would result in a 99\,\% agreement between models and
observations. We further have to consider the possibility that star formation,
and in particular the mass accretion process, might take such a diversity of
paths that no simple unifying model can expect to capture them all, leading to
low agreements even for future much improved observational samples and
models.

\section{Conclusions}
\label{conclusions}
 
We extracted mass accretion rates within cores generated by gravoturbulent
simulations and inserted them into a simple protostellar evolutionary scheme to
calculate protostellar evolutionary tracks. A principle dependence of the
position of these tracks in the T$_{\rm bol}$-L$_{\rm bol}$ diagram on the final
mass is found. A detailed analysis, however, shows that we are not able to
determine the final mass of a particular object from its measured bolometric
temperature and luminosity more accurately than a factor of two. The particular
shape of an evolutionary track is largely determined by the accretion history
and by its final mass. Hence, in the context of the gravoturbulent scheme,
unique evolutionary tracks do not exist in the Class\,0 or Class\,0/1 phase, as
opposed to the ensuing pre-main sequence evolution. It also implies that T$_{\rm
bol}$ is not always a reliable guide to the envelope-protostar mass ratio, as
suggested also by Young \& Evans \cite{2005ApJ...627..293Y}.

A 3D-KS-test was used to compare the distribution of our evolutionary tracks in
the T$_{\rm bol}$-L$_{\rm bol}$-M$_{\rm env}$ parameter space with an
observational sample of Class\,0 objects. By varying free parameters associated
with the evolutionary scheme, we are able to determine constraints for some of
the parameters ($T_{\rm env}$, $frac_{\rm env}$, $M_{\rm extra}$, $M_{\rm eff}$,
$p$). Others were found to have no significant influence on the agreement
between models and observations ($t_0$, $\alpha$, $R_{\rm in}$, $\kappa$). A
comparison of the parameter values obtained by our method with observational
constraints from individual objects (e.g. for $frac_{\rm env}$ or $p$) can be
conducted. This supports the reliability of our method as similar parameter
values are obtained.

Only gravoturbulent models generating model stars with a specific density at the
start of the collapse and Class\,0 lifetimes between 2 and
6\,$\cdot$\,10$^4$\,yrs lead to a good agreement with the observations. This
Class\,0 lifetime is consistent with estimates of a few 10$^4$\,yrs based
on number ratios or dynamical timescales of outflows (see e.g. Andr\'e at al.
\cite{2000prpl.conf...59A}). It is shown that the Class\,0 lifetime is
correlated with the local density at the onset of the collapse. The determined
density range agrees well with density measurements for pre-stellar cores.

Gravoturbulent models with $0.5 \le {\mathcal M} \le 3$ lead to the best
agreement with the observations. The failure of models with ${\mathcal M} \ge 6$
to explain the data indicates that the initial conditions influence the
observational properties of Class\,0 sources. The initial mass function appears
not to influence the agreement systematically. The highest probability  that our
distributions of model tracks and observational data points are drawn from the
same basic population is 70\,\%. This is reasonable given the uncertainties in
the observations and the simple assumptions in the models. However, the method
can be readily adapted to compare larger future source samples, different sets
of accretion rates or other envelope models. 

Applying the KS-test to both the `normal' and underluminous, `Taurus-like'
objects in the source sample, we find that the determined probability $P_{\rm
3D}$ is about twice as large when testing the `normal' objects. This is
consistent with the proposal that turbulence governs the accretion rates in the
majority of the objects. 

\section*{Acknowledgements}

We thank A.\,Scholz for tirelessly answering questions about the 3D-KS method.
D.\,Froebrich received financial support from the Cosmo-Grid project, funded by
the Program for Research in Third Level Institutions under the National
Development Plan and with assistance from the European Regional Development
Fund. S.\,Schmeja and R.S.\,Klessen acknowledge support by the Emmy Noether
Programme of the Deutsche Forschungsgemeinschaft (grant no. KL1358/1). Research
at Armagh Observatory is funded by the Department of Culture, Arts and Leisure,
Northern Ireland.

\begin{appendix}

\section{Gravoturbulent Models}
\label{details_gtmodels}

Our simulations describe the evolution of (1) two globally unstable model
clouds that contract from Gaussian initial conditions without turbulence and
(2) 22 models where turbulence is maintained with constant root mean square
Mach numbers. The models are labelled G1/G2 for the Gaussian, and M$\cal M$k$k$
(with rms Mach number $\cal M$ and wavenumber $k$) for the turbulent models,
following SK04. The designators a, b, c distinguish models that have the same
$\cal M$ and $k$ values, but different random realisations of the turbulent
driving field; they also differ in the time when self-gravity is `switched
on'. 

The gt-models are computed in normalised units. To scale to physical units, we
adopt a mean density of $n({\rm H}_2) = 10^5\,$cm$^{-3}$ and a temperature of
11.3\,K corresponding to a sound speed $c_{\rm s} = 0.2\,{\rm km\,s}^{-1}$. For
the two Gaussian models, the total mass present is $220\,{\rm M}_{\odot}$ and
the size of the cube is $0.34\,$pc. For the turbulent models, the cube contains
a mass of $120\,{\rm M}_{\odot}$ within a volume of ($0.28\,{\rm pc})^3$. The
mean thermal Jeans mass in all models is $\langle M_{\rm J} \rangle =
1\,$M$_{\odot}$ and the global free-fall timescale is $\tau_{\rm ff} =
10^5\,$yr. Requiring that the local Jeans mass is always resolved by at least
100 gas particles (Bate \& Burkert \cite{1997MNRAS.288.1060B}), the resolution
limit is $0.058\,$M$_{\odot}$ in all our turbulent models and
$0.044\,$M$_{\odot}$ in the Gaussian models.

The radius of a sink particle is fixed at 280\,AU. Infalling gas particles
undergo several tests to check if they remain bound to the sink particle before
they are considered accreted. We cannot resolve the evolution in the interior
of the control volume defined by the sink particle. Because of angular momentum
conservation most of the infalling matter will accumulate in a protostellar
disc within which it is transported inwards by viscous and possibly
gravitational torques (Pringle \cite{1981ARA&A..19..137P}; Papaloizou \& Lin
\cite{1995ARA&A..33..505P}; see also Jappsen \& Klessen
\cite{2004A&A...423....1J} for the models discussed here). The latter will be
provided by spiral density waves that develop when the disc becomes too
massive. This occurs when mass is loaded onto the disc faster than it is
removed by viscous transport. Altogether, the disc will not prevent or delay
material from accreting onto the protostar for long. It acts as a buffer and
smoothes possible accretion spikes. For the mass range considered here feedback
effects are too weak to halt or delay accretion (Wuchterl \& Klessen
\cite{2001ApJ...560L.185W}; Wuchterl \& Tscharnuter
\cite{2003A&A...398.1081W}). With typical disc sizes of the order of a few
hundred AU, the control volume encloses both protostar and disc. The measured
core accretion rates are hence good estimates of the actual stellar accretion
rates. Strong deviations may be expected only if the protostellar core
fragments into a binary system, in which case the infalling mass is distributed
over two stars. In addition, if material with very high angular momentum is
accreted, a certain mass fraction may end up in a circumbinary disc and not
accrete onto the star at all.

The timescale over which accretion is smoothed by the disc can be approximated
by the viscous timescale
\begin{equation}
t_{\rm v} \approx \alpha_{\rm v}^{-1} (R/H)^2 t_\Phi
\end{equation} 
(Pringle \cite{1981ARA&A..19..137P}), with the viscosity parameter $\alpha_{\rm
v}$, the disc radius $R$, the disc scale height $H$, and the rotational period
$t_\Phi = \Omega^{-1}$. Using typical values of $\alpha_{\rm v} = 0.01$, $R/H =
10$ and $t_\Phi = 1$\,yr this yields a rough estimate of $t_{\rm v} \approx
10^4$\,yrs for the viscous timescale at a radius of 1\,AU. A large fraction of
mass will lie external to this radius, leading to a larger $t_{\rm v}$, but
this will be compensated for by stronger gravitational torques, which can lead
to significantly larger values of the effective viscosity in the disc (see
e.g. Laughlin \& R\'o\.zyczka \cite{1996ApJ...456..279L}). Since we cannot
model the detailed behaviour inside the disc, we take 10$^4$ yrs as a rough
measure for the smoothing timescale. This value is used to smooth the
individual mass accretion rates from the gt-models. Our results do not
significantly depend on the smoothing scale that we apply. Adopting viscous
timescales of 5 or 15$\cdot$10$^3$\,yrs yields very similar results.

There are further smoothing effects and time-shifts, e.g. the luminosity
generated in the centre typically escapes a Class\,0 envelope in about 300\,yrs
(as determined by a random  walk model). Hence, this time lag is neglected in
comparison to the disc viscosity timescale. The viscosity of the disc has
additionally the effect that material transported from the envelope onto the
disc needs some time before it gets accreted onto the star and the accretion
luminosity is generated. Hence, the observed envelope masses and luminosities
do not represent the same time in the evolution. The observed luminosity
corresponds to that of an envelope mass which was present about a viscosity
timescale earlier. However, since this effect does not become significant until
the disc is formed, it can be neglected in the very early stages of the
evolution. In later stages, the evolution along the T$_{\rm bol}$-L$_{\rm bol}$
track is slower and the luminosity only slightly changes within one viscosity
timescale. In particular the change is smaller than typical errors of the
measurements. Hence, we also neglect this effect in our calculations.

\section{Evolutionary Model}
\label{details_emod}

Accretion rates applied in the e-model to date assume specific functional forms
with a fixed time scale for all protostars. As a consequence, even without a
statistical comparison, it fails to explain the data.  To overcome this, new
parameters were introduced by Myers et al. \cite{1998ApJ...492..703M}. They
took different evolutionary timescales for the envelope and the accretion, and
subsequently found that the observations required (1) the accretion time scale
to exceed the envelope timescale by a factor of a few and (2) the final mass of
the star to be quite a small fraction of the original envelope mass. 

In addition, there are several free parameters in the e-model. Some of their
values are not as yet tightly constrained through observations or theory.
Plausible values, however, predict evolutionary tracks across the same region
as occupied by the observed protostars (e.g. Froebrich et al.
\cite{2003MNRAS.346..163F}). Nevertheless, an immediate problem encountered
with the basic model was that the protostars spent little time in the early
Class\,0 phase but lingered in the late Class\,0 phase for periods exceeding
10$^5$\,yr, implying contrary statistics to those observed. To overcome this, a
distinct mass component was introduced into the envelope. In the e-model, this
arbitrary mass would not accrete, but would be dispersed directly. The
rationale was that star formation remains inefficient due to outflow activity
from the protostar. The second arbitrary parameter was the fraction of mass
contained within a flattened distribution. That is, it is assumed from the
outset, that a small fraction of the gas is available to accrete but does not
add to the obscuration of the protostar. However, given these additions, it
becomes difficult to relate the predictions uniquely to the chosen parameter
set through the KS-test if we attempt to calculate analytically which parameter
values would lead to the highest agreement between models and observations.

Additional free parameters in the e-model are an inner density, and inner and
outer radius for the envelope. We have generalised the previous studies by
introducing a power-law index ($p$) for the radial density distribution, where
$\log \rho \propto -p \times \log R$, rather than fixing the value at 1.5.
Instead of the inner density we employ the total envelope mass as the variable.
In addition, rather than the outer radius, we adopt the outer temperature
$T_{\rm env}$ as the variable which then determines the outer radius. Most of
the mass initially contained in the envelope will eventually fall onto the
protostar. The accretion luminosity depends on the collapse radius and the
contraction of the central hydrostatic object. Here, we assume that a constant
density core develops.

\subsection*{Equations}

In the following we describe in more detail the equations and parameters
used in the evolutionary model for this work.

The initial mass of the envelope and circumstellar disk, $M_0$, is found by 
integrating the mass accretion rate, $\dot M_{\rm a}(t)$, over time. Hence, the 
envelope/disk mass is given by
\begin{equation}
   M_{\rm e,d}(t) = M_0 - \int\limits_0^t \dot M_{\rm a}(t') \cdot dt'.
\end{equation}
The mass just in the spherical envelope is taken as
\begin{equation}
   M_{\rm env}(t)= frac_{\rm env} \cdot M_{\rm e,d}(t).
\end{equation}
In the first approximation, the accreting material is taken as instantaneously
reaching the central hydrostatic core, defining the protostellar radius, $R_*$.
A fraction ($1\,-\,\eta(t)$) of this material is accreted onto the star i.e.
\begin{equation}
   M_*(t) =  \int\limits_0^t  (1\,-\,\eta(t')) \cdot \dot M_{\rm a}(t') \cdot dt'.
\end{equation}
The parameter $\eta(t)$ will depend on the jet launching mechanism. We suppose 
that it is a function of the relative accretion rate i.e.
\begin{equation}
 \eta(t) = M_{\rm eff} \cdot \left[\frac{\dot M_{\rm a}(t)}{\dot M_{\rm
 max}}\right]^\zeta 
\label{eqn-hm}
\end{equation}
where $\dot M_{\rm max}$ is the peak accretion rate and $\zeta = 2$ has been
assumed to date. Thus, $M_{\rm eff}$ is the maximum fraction of material ejected
into the jets.

We assume here that the hydrostatic core grows at constant density during the
very early stages of star formation (no energy released through contraction): 
\begin{equation}
   \frac{R_*(t)}{R_\odot} = 4.24 \cdot 
   \left[ \frac{M_*(t)}{M_\odot} \right]^{1/3} \cdot 
   \left[ \frac{M_0}{M_\odot} \right]^{2/3}.
\end{equation}
We have found that alternative treatments produce similar results. The
particular dependence on envelope mass corresponds to the case where the final
escape speed of the jets is constant. 

The gravitational energy released is assumed to end up in radiation or in the
jets. We thus write the accretion luminosity as
\begin{equation}
 L_{\rm acc}(t)  = (1 - \eta(t)) \cdot G \cdot {\dot M_{\rm a}(t)} \cdot M_*(t)
 / R_*(t). 
\end{equation}

The initial {\em cloud core} mass could exceed the {\em envelope} mass. That
is, some extra mass is introduced. This mass is dispersed rather than being
accreted. In other words, we can check to see if there is some form of feedback
or turbulent dissipation in operation which maintains a cooler core during the
early stages. We assume a form
\begin{equation}
   M_{\rm core}(t) =  M_{\rm env}(t) \cdot \left[1 + M_{\rm extra} \cdot 
   \left(\frac{t+t_0}{t_0}\right)^{-2\alpha}\right],
\label{eq_mextra}
\end{equation}
where $\alpha$ and $t_0$ describe the dissipation of $M_{\rm extra}$. Note that
for $M_{\rm extra}$\,=\,0 the particular values for $t_0$ and $\alpha$ are
superfluous and $M_{\rm core}(t)$ is equal to $M_{\rm env}(t)$.

The radius corresponding to the outer edge of the envelope is determined by an 
assumed outer temperature, $T_{\rm env}$:
\begin{equation}
   R_{\rm env}(t) = \frac{1}{T_{\rm env}^2} \cdot \left[\frac{L_{\rm acc}(t)}{2 \pi
   \cdot \sigma}\right]^{1/2},   
\end{equation}
where $\sigma$ is the Stefan-Boltzmann constant.

Given an inner radius for the envelope, $R_{\rm in}$ (possibly fixed by the
centrifugal barrier condition) and a power-law radial distribution of density
with index $-p$, the inner density of the envelope can be written
\begin{equation}
   \rho_{\rm in}(t) = \frac{(3-p) \cdot M_{\rm core}(t)}{4 \pi \cdot R_{\rm in}^3 \cdot
   \varepsilon(t)} 
\end{equation}
where
\begin{equation}
   \varepsilon(t) = \left( \frac{R_{\rm env}(t)}{R_{\rm in}}\right)^{3-p} - 1.
\end{equation}
The inner temperature of the envelope is
\begin{equation}
    T_{\rm in}(t) = \left[\frac{L_{\rm acc}(t)}{2 \pi \cdot \sigma \cdot
    R_{\rm in}^2}\right]^{1/4}. 
\end{equation}
This yields an optical depth through the envelope proportional to the
emissivity: 
\begin{equation}
   \tau_{\rm e}(t) = \kappa \cdot \rho_{\rm in}(t) \cdot R_{\rm in} \cdot \frac{1 -
   \varepsilon(t)^{1 - p}}{p - 1}. 
\end{equation} 
Following Myers et al. \cite{1998ApJ...492..703M}, we take $\kappa$ as the
emissivity at 12\,$\mu$m, $A = 1.59~10^{-13}$~cm$^2$\,g$^{-1}$\,Hz$^{-1}$, $h$
as the Planck constant, $k$ the Boltzmann constant, and calculate the bolometric
temperature:
\begin{equation}
   T_{\rm bol}(t) = \frac{\Gamma(9/2)\cdot \zeta(9/2)}{\Gamma(5)\cdot \zeta(5)}
   \cdot \left[\frac{h \cdot \kappa \cdot T_{\rm in}(t)}{k \cdot A \cdot
   \tau_{\rm e}(t)}\right]^{1/2}. 
\end{equation}

\section{3D KS-Test and probabilities}
\label{details_kstest}

To evaluate how well observations and models match, we need to determine the
KS-parameter $D_{\rm 3D}$. First, we note that each data point (T$_{\rm
bol}^{\rm i}$, L$_{\rm bol}^{\rm i}$, M$_{\rm env}^{\rm i}$) can be used to
dissect the parameter space into eight octants. The difference $D_{\rm i}$
between the fraction of model data points and observational data points in each
of the eight octants in the T$_{\rm bol}$-L$_{\rm bol}$-M$_{\rm env}$ parameter
space is calculated for each data point. Here, the index $i$ covers {\it both
the observational and model data} points. In case of $N$ data points, this
procedure results in $8N$ numbers $D_{\rm i}$. The maximum absolute difference
value is then defined as the KS-parameter $D_{\rm 3D} = {\rm MAX}(|{D_{\rm
i}}|)$. Note that this value can range from zero to one.

As described in Sect.\,\ref{obsdata} and \ref{gravomod}, the restrictions
in the source sample and models lead to the problem that we have to compare two
samples with extremely different sizes. The observational sample consists of 27
points. Depending on the set of accretion rates we use, the model sample
consists of about ten thousand points (100 tracks of individual stars and on
average 100 points per single track, considering our timestep of 300\,yrs and
assuming a lifetime of 3$\cdot$10$^4$\,yrs for Class\,0 sources). Hence, the
test described above has to be performed about 10$^4$ times to determine $D_{\rm
3D}$ for one case, resulting in a huge amount of computations. Since we are only
interested in the maximum of the absolute differences between the relative
numbers of the model points and observational points in the octants, we can
constrain the investigated area  to a cuboid containing all the observational
datapoints. A test outside these boundaries will not lead to absolute
differences larger than obtained inside the cuboid. In other words, ${\rm
MAX}(|{D_i}|)$ corresponds to a point situated inside or at the boundary of the
cuboid defined as: [T$_{\rm bol}^{\rm obs, min}$\,$<$\,T$_{\rm
bol}$\,$<$\,T$_{\rm bol}^{\rm obs, max}$; L$_{\rm bol}^{\rm obs,
min}$\,$<$\,L$_{\rm bol}$\,$<$\,L$_{\rm bol}^{\rm obs, max}$; M$_{\rm env}^{\rm
obs, min}$\,$<$\,M$_{\rm env}$\,$<$\,M$_{\rm env}^{\rm obs, max}$]. However,
since most of the model points are still situated within this cuboid, the
required amount of computations is not significantly reduced.

Hence, we do not perform the test at all data points within this area, but
rather at a certain number of points inside the cuboid. We performed extensive
tests in order to establish the best compromise between computing time (number
of points where the test has to be performed) and the accuracy of the method.
As a result, we found that a $5 \times 5 \times 5$ point grid in the cuboid is
sufficient to determine $D_{\rm 3D}$ with the same accuracy as the probability
of the agreement between observations and models (see below). The ranges in
T$_{\rm bol}$, L$_{\rm bol}$, and M$_{\rm env}$ are thus divided into four
equally sized regions. While for T$_{\rm bol}$ linear spacing was chosen,
L$_{\rm bol}$ and M$_{\rm env}$ are divided into equal spaces in logarithmic
units.

In order to convert the $D_{\rm 3D}$ value into a probability of agreement
$P_{\rm 3D}$, a Monte Carlo method has to be applied. For this purpose, we
generated artificial observational data points by randomly selecting 27 model
points out of all model points and treating them as observational data. Since
the observational data is limited to T$_{\rm bol} < 80$\,K, we applied this
limit for the random selections as well. Then the KS-test was performed (using
the same grid positions as with the real observational data points), leading to
a value $D_{\rm 3D}^{\rm MC}$. Repeating this process, a distribution N($D_{\rm
3D}^{\rm MC}$) is created. The probability $P_{\rm 3D}$ can then be determined
by:

\begin{equation}
P_{\rm 3D} = \int\limits_{D_{\rm 3D}}^{1} {\rm N}(D_{\rm 3D}^{\rm MC}) \left/
\int\limits_{0}^{1} {\rm N}(D_{\rm 3D}^{\rm MC}) \right.
\end{equation}
 
All $D_{\rm 3D}$-values need to be converted to $P_{\rm 3D}$ by such a
Monte Carlo simulation. To minimise the computational needs we tested if this is
needed for every single $D_{\rm 3D}$-value. It turned out that the distributions
N($D^{\rm MC}_{\rm 3D}$) obtained by a Monte Carlo simulation for one gt-model
but different parameter sets in the e-model are similar. Hence all $D_{\rm
3D}$-values of one gt-model can be converted into $P_{\rm 3D}$ using the same
distribution. However, distributions N($D^{\rm MC}_{\rm 3D}$) for distinct
gt-models are partly very different. Thus, to find the best fitting model
combination we selected the parameter combination leading to the smallest
$D_{\rm 3D}$-value for each gt-model to determine N($D^{\rm MC}_{\rm 3D}$). This
was then used to calculate the agreement $P_{\rm 3D}$.

The total number of random selections to build up the distribution N($D_{\rm
3D}^{\rm MC}$) was a critical point concerning the accuracy of the method.
Hence, we tested for which number of repeats the final values of $P_{\rm 3D}$
did not vary by more than one percent if the same test was repeated. For our
distribution of model points, this was the case for 3$\cdot$10$^4$ random
selection processes. This accuracy is not degraded by the $5 \times 5 \times 5$
grid, chosen for performing the KS-test, instead of doing the test at all
datapoints in the cuboid. 

There are nine free parameters in the e-models which we kept variable. Our task
here is to find the combination of these parameters that leads to the best
agreement with the models. A test of all possible combinations with an accurate
spacing requires an immense amount of computations. We thus decided to use a
Monte Carlo approach for this task as well. We randomly selected 100 values for
each parameter from the range given in Table\,\ref{params} and determined
$D_{\rm 3D}$ for these 100 random combinations. We then selected the parameter
sets that lead to $D_{\rm 3D}$-values which were smaller than $(D^{\rm max}_{\rm
3D} + D^{\rm min}_{\rm 3D}) / 2$. The maximum and minimum for each parameter
value used in these selected sets was then taken to re-set the ranges out of
which parameter values are selected. Then again 100 e-models with randomly
selected parameters out of the new range were tested. This procedure was
repeated 30 times, finally leading to 3100 tested e-models. These models are
used to determine the best fitting parameter ranges (Table\,\ref{best_agree})
for the individual gt-models. These ranges were then used to determine another
10000 random parameter combinations, in order to find the best agreement.

We tested for a subset of parameters and smaller ranges if the Monte Carlo
method leads to the same results as a straight forward testing of all possible
parameter combinations. Both, the smallest $D_{\rm 3D}$-value and the best
fitting ranges for the parameter values, are found to be similar. Note that the
smallest found $D_{\rm 3D}$-value has to be considered an upper limit, since not
the whole parameter space was tested. Alternatively the probability $P_{\rm 3D}$
is a lower limit.

\end{appendix}

\label{lastpage}

\end{document}